\documentclass{emulateapj}

\shorttitle{Redshift Distribution of Extragalactic 24 $\micron$ Sources}
\shortauthors{Desai et~al.}

\newcommand\bootes{Bo\"{o}tes}

\submitted{Accepted for publication in ApJ}

\begin{document}

\title{Redshift Distribution of Extragalactic 24 \boldmath$\micron$\unboldmath{} Sources\altaffilmark{1,2}}

\author{Vandana~Desai\altaffilmark{3,4}, B.~T.~Soifer\altaffilmark{3,4}, Arjun~Dey\altaffilmark{5}, Buell~T.~Jannuzi\altaffilmark{5}, Emeric~Le~Floc'h\altaffilmark{6,7,8}, Chao~Bian\altaffilmark{3}, Kate~Brand\altaffilmark{5,9}, Michael~J.~I.~Brown\altaffilmark{10}, Lee~Armus\altaffilmark{4}, Dan~W.~Weedman\altaffilmark{11}, Richard~Cool\altaffilmark{7}, Daniel~Stern\altaffilmark{12}, Mark~Brodwin\altaffilmark{5}}

\altaffiltext{1}{Based on observations made with the \textit{Spitzer Space Telescope}, operated by the Jet Propulsion Laboratory under NASA contract 1407.}
\altaffiltext{2}{Some of the data presented herein were obtained at the W.~M.~Keck Observatory, which is operated as a scientific partnership among the California Institute of Technology, the University of California, and the National Aeronautics and Space Administration. The Observatory was made possible by the generous financial support of the W.~M.~Keck Foundation.}
\altaffiltext{3}{Division of Physics, Mathematics and Astronomy, California Institute of Technology, Pasadena CA 91125}
\altaffiltext{4}{Spitzer Science Center, California Institute of Technology, Pasadena CA 91125}
\altaffiltext{5}{National Optical Astronomy Observatory, Tucson AZ 85726-6732}
\altaffiltext{6}{Spitzer Fellow}
\altaffiltext{7}{Steward Observatory, University of Arizona, Tucson AZ 85721}
\altaffiltext{8}{Institute for Astronomy, University of Hawaii, Honolulu HI 96822}
\altaffiltext{9}{Space Telescope Science Institute, Baltimore MD 21218}
\altaffiltext{10}{School of Physics, Monash University, Clayton, Victoria 3800, Australia}
\altaffiltext{11}{Astronomy Department, Cornell University, Ithaca NY 14853}
\altaffiltext{12}{Jet Propulsion Laboratory, California Institute of Technology, Pasadena CA 91109}

\begin{abstract}

We present the redshift distribution of a complete, unbiased sample of
24~$\micron$ sources down to $f_{\nu}(24 \micron) = 300$ $\mu$Jy
($5\sigma$).  The sample consists of 591 sources detected in the
\bootes \ field of the NOAO Deep Wide-Field Survey.  We have obtained
optical spectroscopic redshifts for 421 sources (71\%).  These have a
redshift distribution peaking at $z \sim 0.3$, with a possible
additional peak at $z \sim 0.9$, and objects detected out to $z =
4.5$.  The spectra of the remaining 170 (29\%) exhibit no strong
emission lines from which to determine a redshift.  We develop an
algorithm to estimate the redshift distribution of these sources,
based on the assumption that they have emission lines but that these
lines are not observable due to the limited wavelength coverage of our
spectroscopic observations.  The redshift distribution derived from
all 591 sources exhibits an additional peak of extremely luminous
(L$_{8-1000 \micron} > 3 \times 10^{12}$ L$_{\odot}$) objects at $z
\sim 2$, consisting primarily of sources without observable emission
lines.  We use optical line diagnostics and IRAC colors to estimate
that 55\% of the sources within this peak are AGN-dominated.  We
compare our results to published models of the evolution of infrared
luminous galaxies.  The models which best reproduce our observations
predict a large population of star-formation dominated ULIRGs at $z >
1.5$ rather than the AGN-dominated sources we observe.  

\end{abstract}


\section{Introduction}
\label{sec:Introduction}

An important goal in extragalactic astronomy is to understand when,
where, and how the stellar content of the universe formed.  Based on
rest-frame ultraviolet and optical measures, the star formation rate
per comoving volume is approximately constant from $z \approx 6$ to $z
\approx 2$ and then smoothly decreases by a factor of ten until the
present epoch
\citep{Gallego95,Lilly96,Madau96,Connolly97,Treyer98,Flores99,
Steidel99,Wilson02,Giavalisco04}.  There is also increasing evidence
that galaxies with large stellar masses formed most of their stars
earlier than lower-mass galaxies
\citep{Tinsley68,Cowie96,Heavens04,Juneau05}.

Observations of the far-infrared background radiation with the
\textit{Cosmic Background Explorer}
\citep[\textit{COBE};][]{Puget96,Fixsen98,Hauser98,Dwek98,Lagache99,Lagache00,Finkbeiner00}
show that the infrared energy density is comparable to the combined
UV, visible, and near-infrared energy density, indicating that half of
the light originally emitted in the UV and optical has been
intercepted by dust, reprocessed, and re-emitted in the infrared
\citep{Gispert00}.  What kinds of galaxies contribute to the infrared
peak in the extragalactic background light?  The earliest mid-infrared
observations of extragalactic sources revealed a class of galaxies
which emit the bulk of their luminosity in the infrared
\citep{Low68,Kleinmann70b,Kleinmann70a,Rieke72}.  Early observations
with the \textit{Infrared Astronomical Satellite} (\textit{IRAS})
showed that luminous infrared galaxies (LIRGs; $10^{11} < {\rm
L}_{8-1000 \micron} ({\rm L}_{\odot}) < 10^{12}$) and ultraluminous
infrared galaxies (ULIRGs; ${\rm L}_{8-1000 \micron} > 10^{12}$ ${\rm
L}_{\odot}$) do not contribute substantially to the total infrared
emission in the local ($z < 0.05$) universe \citep{Soifer91}.

However, there is strong evidence that infrared-bright populations
become more significant at higher redshift.  This evidence includes
number counts obtained with \textit{IRAS} \citep{Hacking87} and the
\textit{Infrared Space Observatory}
\citep[\textit{ISO};][]{Oliver97,Altieri99,Aussel99,Elbaz99,Gruppioni02,Metcalfe03}.
Spectroscopic follow-up of these sources has directly confirmed the
increasing significance of the infrared-bright population out to $z
\sim 0.3$ \citep{Kim98,Serjeant01}, a redshift set by the sensitivity
limits of the existing mid-infrared surveys.  Progress in constraining
the population of infrared-luminous galaxies at higher redshift has
been made via surveys at submillimeter
\citep[e.g.][]{Smail97b,Chapman01,Blain02} and radio \citep{Cowie04}
wavelengths.  These data provide further evidence of strong evolution
in the infrared-luminous population, but the number of sources probed
are small, and may suffer from selection biases.  For instance,  
spectroscopy of submillimeter sources has relied on accurate positions
obtained from radio counterparts, biasing spectroscopic samples to
sources that are bright in both the submillimeter and the radio
\citep{Chapman05}.  The sensitive Multiband Imaging Photometer for the
\textit{Spitzer Space Telescope} \citep[MIPS;][]{Rieke04,Werner04} has
allowed the detection of many faint infrared sources.  Number counts
at 24~$\micron$ determined by \textit{Spitzer} observations confirm
the strong evolution previously observed in the infrared luminous
population \citep{Chary04,Dole04,Marleau04,Papovich04}.

Although number counts provide valuable information on the evolution
of the infrared population, they do not allow a robust distinction
between luminosity and density evolution.  For this, redshifts are
needed.  The redshift distribution is shaped by the evolution of
galaxy SEDs and luminosity functions, which in turn depend on the
physics of galaxy evolution: star formation, AGN activity, dust
obscuration, mergers, and feedback \citep{Lagache05}.  The redshift
distribution is therefore a useful constraint on the importance of
these processes.

Several attempts have been made to estimate the redshift distribution
of 24~$\micron$ sources using photometric redshifts.  For sources with
$f_{\nu}(24 \micron) > 83$ $\mu$Jy, these studies consistently reveal
a primary peak in the redshift distribution at $z < 1$, with small
differences in the precise location of this peak attributable to
cosmic variance \citep{PerezGonzalez05,LeFloch05,Caputi06,Papovich07}.
Such a peak can be attributed to strong luminosity and/or density
evolution in the infrared population out to $z \sim 1$
\citep{Chary01,Lagache04,Gruppioni05,PerezGonzalez05}.  However,
contradictory results are found for the existence of a predicted
secondary peak at $z \sim 2$ \citep{PerezGonzalez05,Caputi06}.  Such a
secondary peak may be expected from a significant high-redshift
population of either highly absorbed AGN-dominated sources or PAH-rich
starburst-dominated sources.  These would have a local peak in their
SEDs at $\approx$8 $\micron$, due either to continuum adjacent to deep
absorption features or to line emission, respectively.  This 8
$\micron$ peak would enhance the detectability of these sources at $z
\approx 2$ in 24 $\micron$ surveys.  Uncertainties in the photometric
redshifts may be responsible for the contradictory observational
results regarding the existence of this peak, since previous studies
have few ($\la$25) spectroscopic redshifts above $z = 1.5$.
Photometric estimates of the redshift distribution for more shallow
surveys ($f_{\nu}(24 \micron) \gtrsim 200$ $\mu$Jy) also produce mixed
results for the $z \sim 2$ peak \citep{RowanRobinson05,Babbedge06}.
Recent spectroscopic surveys of 24~$\micron$ sources are complete only
to about the same redshift ($z \sim 1$) reliably probed by photometric
redshifts \citep{Papovich06,Marleau07}.

In this work, we present new measurements of the redshift distribution
of 24~$\micron$ sources down to 300 $\mu$Jy.  Our sample is selected
from 23 noncontiguous areas spread across 8.74 deg$^2$ of the 24
$\micron$ map of the \bootes \ field of the NOAO Deep Wide-Field
Survey \citep[NDWFS;][]{Jannuzi99}.  Our results are therefore
minimally affected by variations in the density of the 24~$\micron$
source population due to structures on the scale of $\sim$0.5 deg$^2$.
Although our sample is modest (591 sources; comparable in size to the
smaller studies mentioned above), a larger fraction (71\%) have
spectroscopic redshifts.  The remaining 29\% exhibit no spectral
features from which to determine a redshift.  Given the limitations of
photometric redshifts for 24~$\micron$ sources at $z > 1$, we have
developed an algorithm for estimating the redshift distribution of
this population from their continuum spectra.  Briefly, this described
(algorithm in detail in \S{\ref{sec:GalaxiesWithoutEmissionLines}}) is
based on the assumptions that bright 24 $\micron$ sources are
undergoing strong star formation and/or AGN activity, and the emission
lines resulting from this activity are able to penetrate the
surrounding dust.  These assumptions hold true for $\approx$70\% of
the 24 $\micron$ sources in our sample, and also hold for very
obscured local ultraluminous infrared sources
\citep[e.g.][]{Sanders88,Kleinmann88,Armus89,Allen91,Cutri94,Veilleux95,Veilleux97,Duc97}.
Under these assumptions, the lack of observed emission lines for
certain sources is due to the limited wavelength coverage of our
spectroscopic observations.

We describe our targeting strategy in \S \ref{sec:Targeting}, our
targeting completeness in \S \ref{sec:completeness}, and our
spectroscopic observations in \S \ref{sec:SpectroscopicObservations}.
Our results are presented in \S{\ref{sec:Results}}.  In particular, we
present the redshift distribution of emission-line galaxies
(\S{\ref{sec:GalaxiesWithEmissionLines}}), the redshift distribution
of galaxies that exhibit no emission lines
(\S{\ref{sec:GalaxiesWithoutEmissionLines}}), and the redshift
distribution for the entire sample
(\S{\ref{sec:RedshiftDistribution}}).  In \S{\ref{sec:Discussion}}, we
discuss the shape of the overall redshift distribution and compare our
observations to existing models of the evolution of infrared galaxies.
Finally, we summarize our conclusions in \S{\ref{sec:Conclusions}}.
All calculations are carried out assuming a cosmology with H$_0 = 70$
km s$^{-1}$ Mpc$^{-1}$, $\Omega_0 = 0.3$, and $\Omega_{\Lambda} =
0.7$.

\section{Targeting Strategy}
\label{sec:Targeting}

We have conducted an optical spectroscopic survey of $\sim$600 24
$\micron$ sources in order to determine their redshift distribution.
The spectroscopic targets were drawn from a catalog of $\sim$22,000
sources detected in a MIPS survey of 8.74 deg$^2$ of the \bootes \
field of the NDWFS.  The 5$\sigma$ point-source depth of the imaging
is 300 $\mu$Jy, and this sets the flux limit of both the detection
catalog and our spectroscopic survey.

\begin{figure}
\plotone{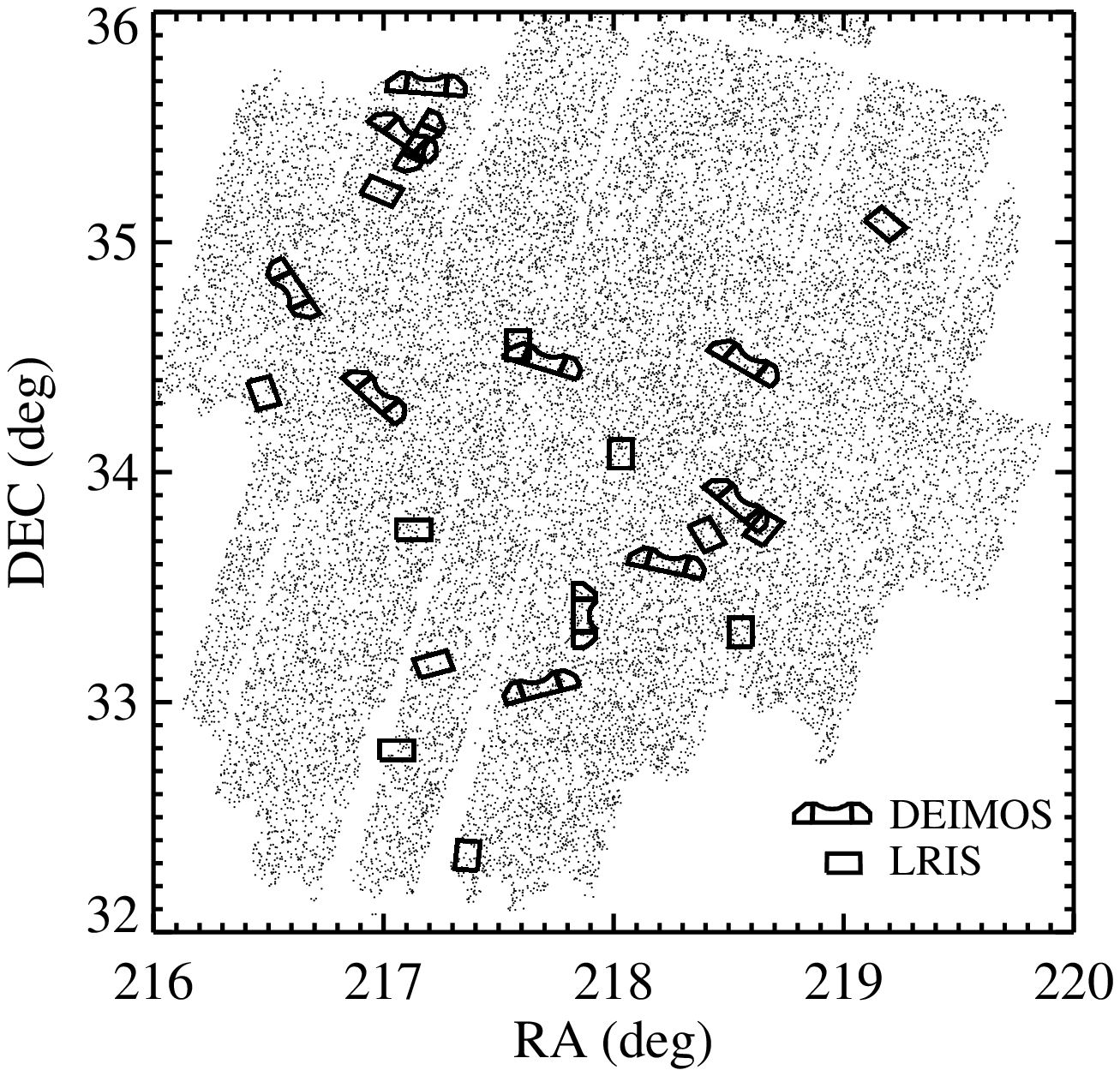}
\caption{Schematic of Keck spectroscopic survey.  Points indicate the
locations of all \bootes \ sources with $f_{\nu}(24 \micron) \ge 300$
$\mu$Jy, the sample from which our optical spectroscopic targets were
drawn.  Overlaid are the positions of the LRIS and DEIMOS masks that
comprise the spectroscopic survey.  Mask positions were chosen to
target rare sources with the most extreme infrared-to-optical flux
density ratios.  The final redshift distribution is corrected for this
selection as described in \S{\ref{sec:Targeting}}.  The total area of
the spectroscopic survey is 0.33 deg$^2$.}
\label{fig:schematic}
\end{figure}

The spectroscopy was carried out using the Deep Imaging Multi-Object
Spectrograph \citep[DEIMOS;][]{Faber03} and the Low Resolution Imaging
Spectrometer \citep[LRIS;][]{Oke95}, both on the W.~M.~Keck 10-meter
telescopes.  In total, we observed 11 DEIMOS and 12 LRIS masks.  Mask
positions were chosen to maximize the number of high-priority targets,
which are defined to be sources with $f_{\nu}(24 \micron) > 750$
$\mu$Jy and faint optical magnitudes ($R > 24$ mag or $f_{\nu}(R) <
0.74$ $\mu$Jy).  Figure \ref{fig:schematic} shows the positions of the
masks on the sky.  The remaining targets on each mask were selected
from the sources with $f_{\nu}(24 \micron) \ge 300~\mu$Jy, without
regard to $R-[24]$ color.  To $f_{\nu}(24 \micron) \ge 300$ $\mu$Jy,
24~$\micron$ sources have a surface density of about 2400 deg$^{-2}$.
There are $\approx$50 per 0.02 deg$^2$ DEIMOS mask and $\approx$25 per
0.01 deg$^2$ LRIS mask, resulting in a total of 818 over the 0.33
deg$^2$ combined mask area. We targeted 544, or 67\%.  (Our final
sample includes 47 supplementary sources described in Section
\ref{sec:completeness}.) The untargeted sources were excluded mainly
due to slit collisions.  In the case of slit collisions, we
preferentially targeted faint (but not invisible) optical sources.
Although their combined area is modest, our spectroscopic masks span
8.74 deg$^2$ of the \bootes \ field.  This makes our survey less
subject to the effects of cosmic variance due to structures on the
scale of approximately half a square degree.

In designing the spectroscopic masks, we attempted to position the
slits on the optical ($R$-band) counterparts of the 24 $\micron$
targets.  For reference, the NDWFS $R$-band imaging has a 3$\sigma$
point-source depth of $\sim$26 Vega magnitudes.\footnote{See
http://www.noao.edu/noao/noaodeep/ for more information regarding the
depth and coverage of the NDWFS.}  The position of the optical
counterpart was determined in the following way.  For each 24
$\micron$ source, the SExtractor $R$-band catalog was searched for
nearby sources.  For extended sources, we match detections in
different bands when their centroids are within an ellipse defined by
the second-order moments of the optical light distribution (as defined
by SExtractor). For both compact and extended sources we match
detections in different bands when their centroids are within
1$\arcsec$ of each other. When matching bright sources, where
centroids in different bands can be offset or optical sources are well
resolved, matching MIPS sources to SExtractor catalogs has some
advantages over simple $R$-band aperture photometry at the position of
the MIPS source.  When their are multiple $R$-band detections which
meet our matching criteria, our code opts for the closest centroid.
The plots in this paper show magnitudes computed within a 4$\arcsec$
diameter aperture.  In some cases, we find no $R$-band
counterpart. After visual inspection of these cases, we perform
$R$-band aperture photometry at the MIPS position.  The optical
positions determined by the above method were used as a starting point
for designing the slit masks.  The final slit positions were adjusted
based on visual inspection of the $R$-band and 24 $\micron$ images, as
well as additional imaging \citep{Eisenhardt04} acquired with
\textit{Spitzer's} Infrared Array Camera \citep[IRAC;][]{Fazio04}.
The IRAC images were useful in identifying ambiguous optical
counterparts, since the MIPS targets are all detected by IRAC, which
has a much smaller PSF than MIPS.  In addition, sometimes faint
optical counterparts can be identified by eye even if they did not
appear in the $R$-band SExtractor catalog.

\section{Targeting Completeness}
\label{sec:completeness}

\begin{figure}
\epsscale{0.9}
\plotone{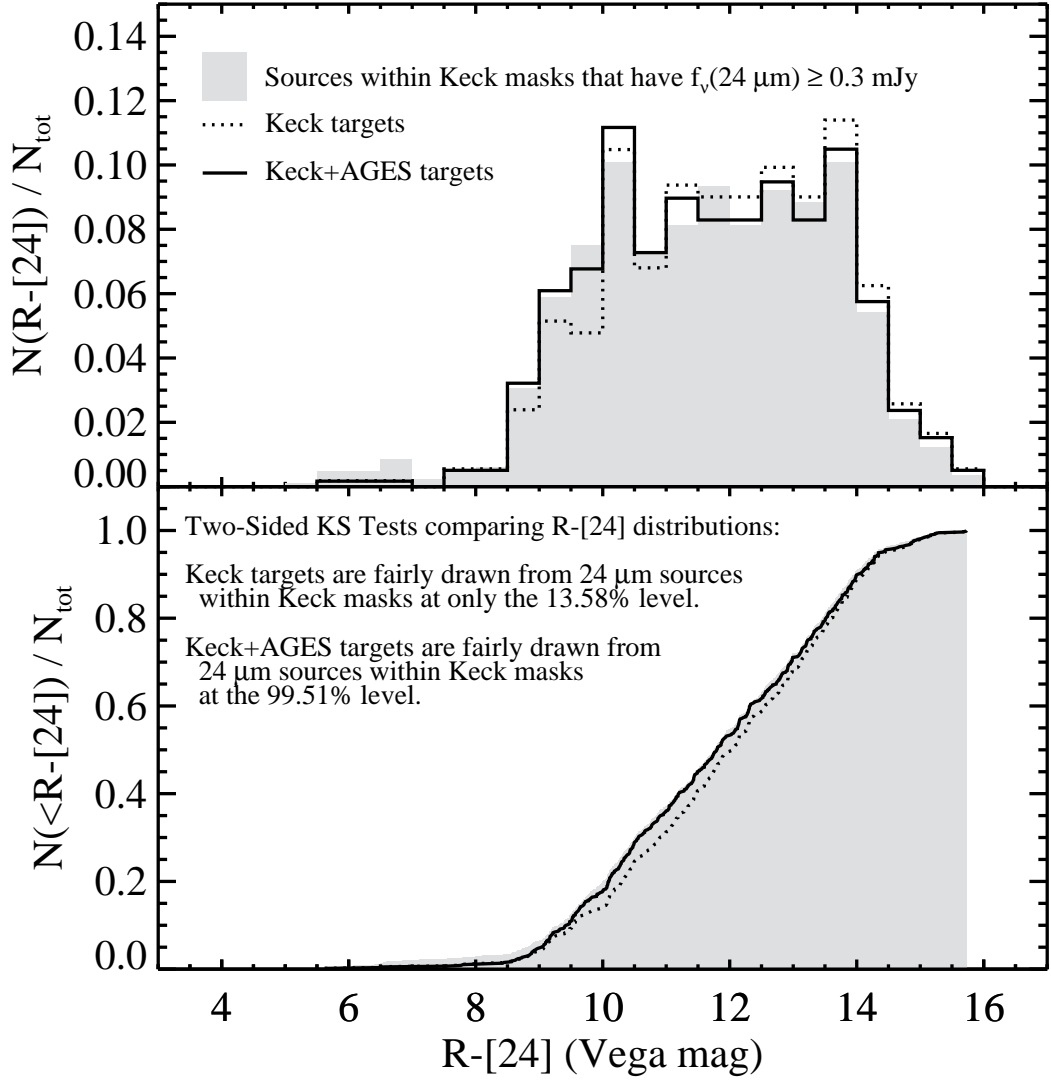}
\caption{Differential (\textit{top}) and cumulative (\textit{bottom})
$R-[24]$ color distributions.  Each of the distributions is normalized
by the number of sources within the represented sample.  Because of
the targeting selection described in \S{\ref{sec:Targeting}}, the Keck
targets include a disproportionately small fraction of sources with
blue $R-[24]$ colors. Kolmogorov-Smirnov (KS) tests show that
supplementing the Keck targets with additional redshifts from AGES
results in a sample that is fairly drawn from the full 24 $\micron$
\bootes \ population in terms of $R-[24]$ color.  KS tests also
confirm that the 24 $\micron$ flux density and $R$-band magnitude
distributions of the Keck+AGES sample are consistent with those of the
full \bootes \ population.}
\label{fig:ColorCompleteness}
\end{figure}

In the following, the term ``Keck spectroscopic targets'' refers to
both the high-priority targets and the flux-limited targets described
above.  The inclusion of high-priority sources as well as the
preference for faint $R$-band sources in cases of slit collisions mean
that our sample of Keck spectroscopic targets is not completely
unbiased.  Figure \ref{fig:ColorCompleteness} shows the differential
and cumulative distributions of $R-[24]$ colors for all of the
$f_{\nu}(24 \micron) \ge 300$ $\mu$Jy sources within the Keck survey
region and for the subset of these that were targeted for
spectroscopy.  A two-sided Kolmogorov-Smirnov (KS) test shows that the
24~$\micron$ flux density distributions of these two sets of galaxies
are indistinguishable.  In contrast, their $R$-band magnitude and
$R-[24]$ color distributions are different at a statistically
significant level, in the sense that our target list includes an
excess of optically faint, red sources.

To correct this bias, we supplement our Keck spectroscopic targets
with 47 optically bright ($R < 23$ mag) sources within the Keck survey
region that were not targeted with LRIS or DEIMOS, but which have
redshifts from the AGN and Galaxy Evolution Survey (AGES; Kochanek et
al.~in preparation).  AGES is a spectroscopic survey of galaxies ($I <
20~{\rm mag}$) and AGN ($I < 21.5$ mag) in the NDWFS \bootes \ field
carried out with the Hectospec instrument \citep{Fabricant05} on the
MMT.  We selected the supplementary AGES sample randomly under the
requirement that each 0.1 $R$-band magnitude bin at $R < 23$ mag have
the same sampling fraction as the average Keck sampling fraction at $R
> 23$ mag (74\%).  The differential and cumulative distributions of
$R-[24]$ colors for the combined Keck and supplementary AGES targets
are shown in Figure \ref{fig:ColorCompleteness}.  Two-sided
Kolmogorov-Smirnov tests confirm that the Keck+AGES sample is fairly
drawn from the population of sources with $f_{\nu}(24 \micron) > 300$
$\mu$Jy that lie within the Keck spectroscopic survey area, in terms
of $f_{\nu}(24 \micron)$, $R$-band magnitude, and $R-[24]$ color.
Thus we conclude that the Keck+AGES targets are an unbiased sample of
sources selected at 24~$\micron$.

\section{Spectroscopic Observations}
\label{sec:SpectroscopicObservations}

The DEIMOS and LRIS optical spectroscopic followup of 24~$\micron$
sources was carried out in May 2004 and May 2005.  The DEIMOS
observations for each mask are composed of 3-4 frames lasting 20 or 30
minutes each, resulting in total exposure times of 90-120 minutes per
mask.  The 600 line mm$^{-1}$ (7500 \AA \ blaze) grating was used with
the GG400 blocking filter, resulting in a mean spectral dispersion of
0.65 \AA \ pixel$^{-1}$.  The central wavelength was set to 7500 \AA.
We typically used 1.0\arcsec \ wide by 10\arcsec \ long slitlets,
corresponding to a FWHM spectral resolution of 4.6 \AA.  Although the
wavelength coverage varies with slit position, a typical DEIMOS
spectrum ranges from 5200-10200 \AA.  We used the DEEP2 data reduction
pipeline\footnote{http://alamoana.keck.hawaii.edu/inst/deimos/pipeline.html}
to perform cosmic ray removal, flat-fielding, co-addition,
sky-subtraction, and wavelength calibration.  The one-dimensional
spectra were extracted using standard IRAF\footnote{IRAF is
distributed by the National Optical Astronomy Observatory, which is
operated by the Association of Universities for Research in Astronomy,
Inc., under cooperative agreement with the National Science
Foundation.} routines \citep{Tody93}.

The LRIS observations are also composed of 3-4 frames lasting 20 or 30
minutes each, yielding total exposure times of 60-120 minutes per
mask.  One mask was observed for a total of only 20 minutes.  The
telescope was offset by a few arcseconds between frames.  The slitlets
were 1\arcsec \ wide by 11\arcsec \ long.  LRIS was configured with
the 5600 \AA \ dichroic.  On the blue side, we used the 400 line
mm$^{-1}$ grism with a central wavelength of 3400 \AA, resulting in a
spectral dispersion of 1.09 \AA \ pixel$^{-1}$ and a FWHM spectral
resolution of 6.9 \AA.  On the red side, we used the 400 line
mm$^{-1}$ grating with a central wavelength of 8500 \AA, resulting in
a spectral dispersion of 1.86 \AA \ pixel$^{-1}$ and a FWHM spectral
resolution of 7.5 \AA.  Our LRIS spectra typically have a wavelength
coverage of 2200-8200 \AA.  We used an IRAF package called
BOGUS\footnote{https://zwolfkinder.jpl.nasa.gov/$\sim$stern/homepage/bogus.html}
to reduce the two-dimensional spectra.  The one-dimensional spectra
were extracted using standard IRAF routines.

Relative spectrophotometric calibration for both the DEIMOS and LRIS
observations was performed using observations of Wolf 1346
\citep{Massey88,Massey90,Oke90}.

\section{Results}
\label{sec:Results}

Our primary observational result is the redshift distribution of 24
$\micron$ sources down to 300 $\mu$Jy.  In the following three
subsections, we present the redshift distributions of the sources with
detected emission lines, of the sources without detected emission
lines, and of the entire sample.

\subsection{Redshift Distribution of Galaxies with Detected Emission Lines}
\label{sec:GalaxiesWithEmissionLines}

\begin{figure}
\epsscale{0.9}
\plotone{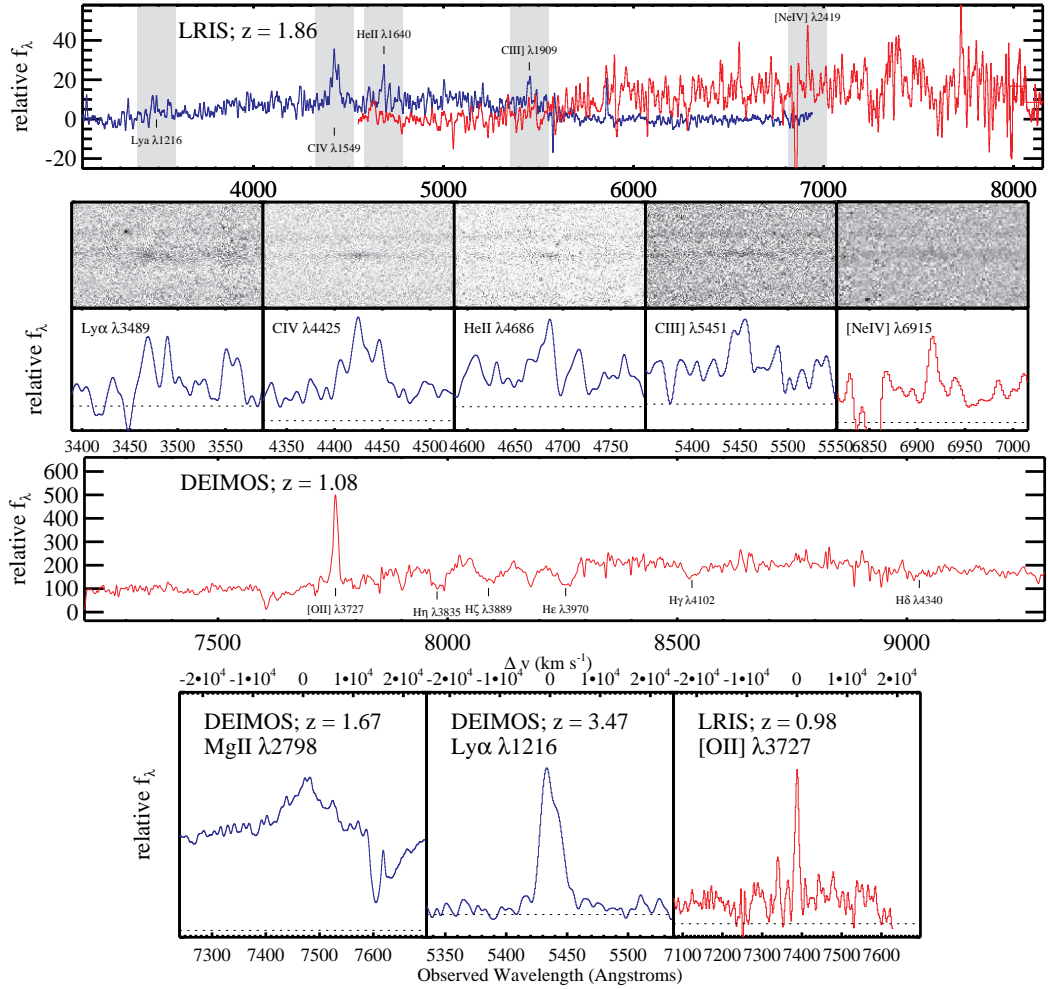}

\caption{Example spectra.  The top row shows an LRIS spectrum
featuring multiple emission lines from which to determine a
redshift. The grey shaded regions indicate regions around the lines
that are shown in expanded form in the second (2D) and third (1D)
rows.  The fourth row shows a DEIMOS spectrum with only one emission
line, but multiple absorption features allow for a robust redshift
determination. In some spectra, only one emission line and no
absorption lines were detected.  In these cases, we identified the
line by its shape, as illustrated in the fifth row.  The zero-level
fluxes are shown as dotted lines for each of the 1D snippets in the
third and fifth rows.}

\label{fig:samplespectra}
\end{figure}

Each of the one- and two-dimensional DEIMOS and LRIS spectra were
visually inspected for spectral lines from which redshifts could be
measured.  Some example spectra are shown in Figure
\ref{fig:samplespectra}, and many more will be provided in a
forthcoming paper (Soifer et al.~in preparation).  Out of the 544
sources that were targeted, 368 spectra (69\%) yielded redshifts.  Two
or more spectral lines were typically available to determine a
redshift.  In 44 cases (12\%), only one spectral line was observed.
Eight of these had absorption features in the Keck spectrum or
additional features from AGES spectra that helped to determine the
redshift.  In the remaining spectra, the single observed emission line
was identified based on its shape and width as Ly$\alpha$ in eight
cases, MgII$\lambda$2798 in three cases, and [OII]$\lambda$3727 in 25
cases.  These single-line sources are included in our sample with
spectroscopic redshifts.

Of the 176 galaxies that we targeted with either DEIMOS or LRIS but
for which we were unable to determine a redshift, six of these were
bright enough in the optical ($R = $ 21.16, 16.58, 19.42, 22.15,
20.12, 18.30 mag) to have redshifts from the AGES Survey.  The AGES
redshifts are $z = $ 1.799, 0.076, 2.230, 1.509, 0.545, and 1.069.
The DEIMOS spectrum of the $z = 0.076$ source did not yield a redshift
because it was saturated.  The LRIS spectrum of the $z = 0.545$ source
showed absorption lines, but identifiable emission lines were lost in
the gap between the red and blue sides.  The differing wavelength
coverage of the Keck spectrographs compared to Hectospec explains the
remaining sources.  The AGES data bring the number of Keck targets
with redshifts up to 374.  Hereafter, we refer to the redshifts of all
Keck targets with either Keck or AGES spectroscopic redshifts as the
Keck sample.  In contrast, the Keck targets supplemented with sources
with bright optical fluxes from AGES, as described in
\S{\ref{sec:completeness}}, will be referred to as the Keck+AGES
sample.

\begin{figure}
\epsscale{0.9} \plotone{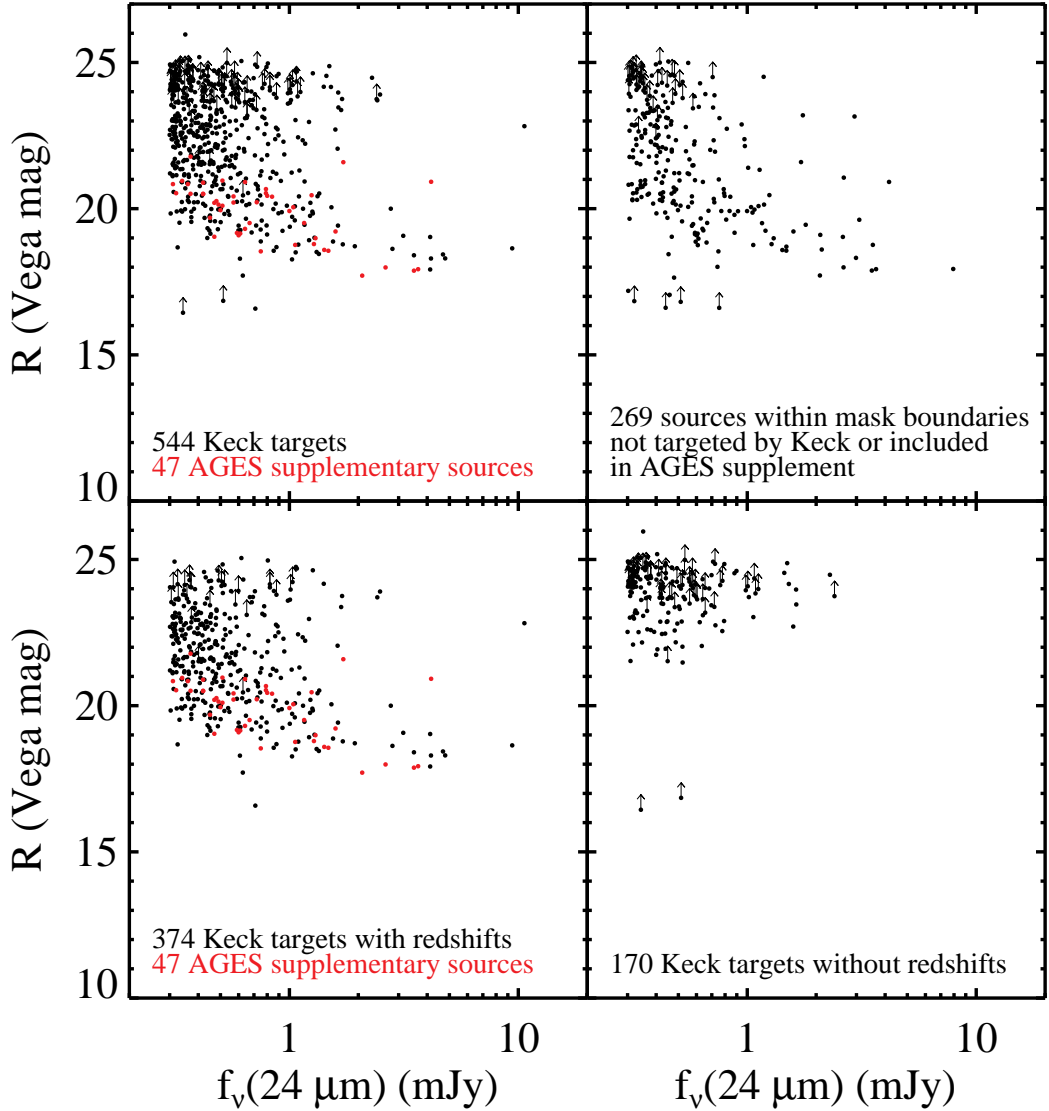}
\caption{$R$-band magnitude versus MIPS 24~$\micron$ flux density for
various subsamples of the 24 $\micron$ population within the Keck
spectroscopic mask boundaries.  This plot illustrates that (1) there
is not a strong correlation between 24 $\micron$ flux density and
$R$-band magnitude; (2) the AGES supplement consists of sources with
$R < 23$ mag that span a large range of $f_{\nu}(24 \micron)$; (3) the
distribution of Keck+AGES targets is similar to that of the untargeted
sources (visual confirmation that the Keck+AGES targets represent a
fair subsample of 24 $\micron$ sources); and (4) the Keck targets
without spectroscopic redshifts tend to have faint $R$-band magnitudes
but a range of $f_{\nu}(24 \micron)$.}
\label{fig:Rversusf24}
\end{figure}

Figure \ref{fig:Rversusf24} shows the $R$-band magnitudes versus 24
$\micron$ flux densities for the Keck+AGES targets, for the sources in
the spectroscopic survey region that were not included in the
Keck+AGES target list, for the Keck+AGES targets with spectroscopic
redshifts, and for the Keck targets without spectroscopic redshifts.
This figure illustrates that the 24~$\micron$ flux density is only
weakly correlated with the $R$-band magnitude, and that the Keck
targets for which we were unable to determine spectroscopic redshifts
are mostly faint in the optical ($R > 21$ Vega mag), but span a range
in 24 $\micron$ flux densities.

\begin{figure}
\includegraphics[scale = 0.75]{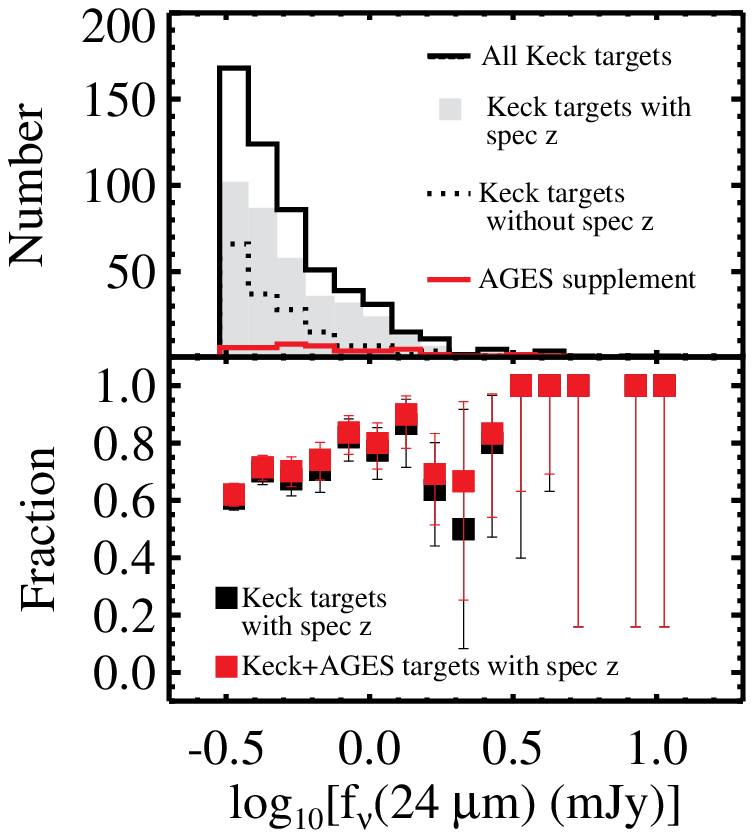} \hfill \includegraphics[scale = 0.75]{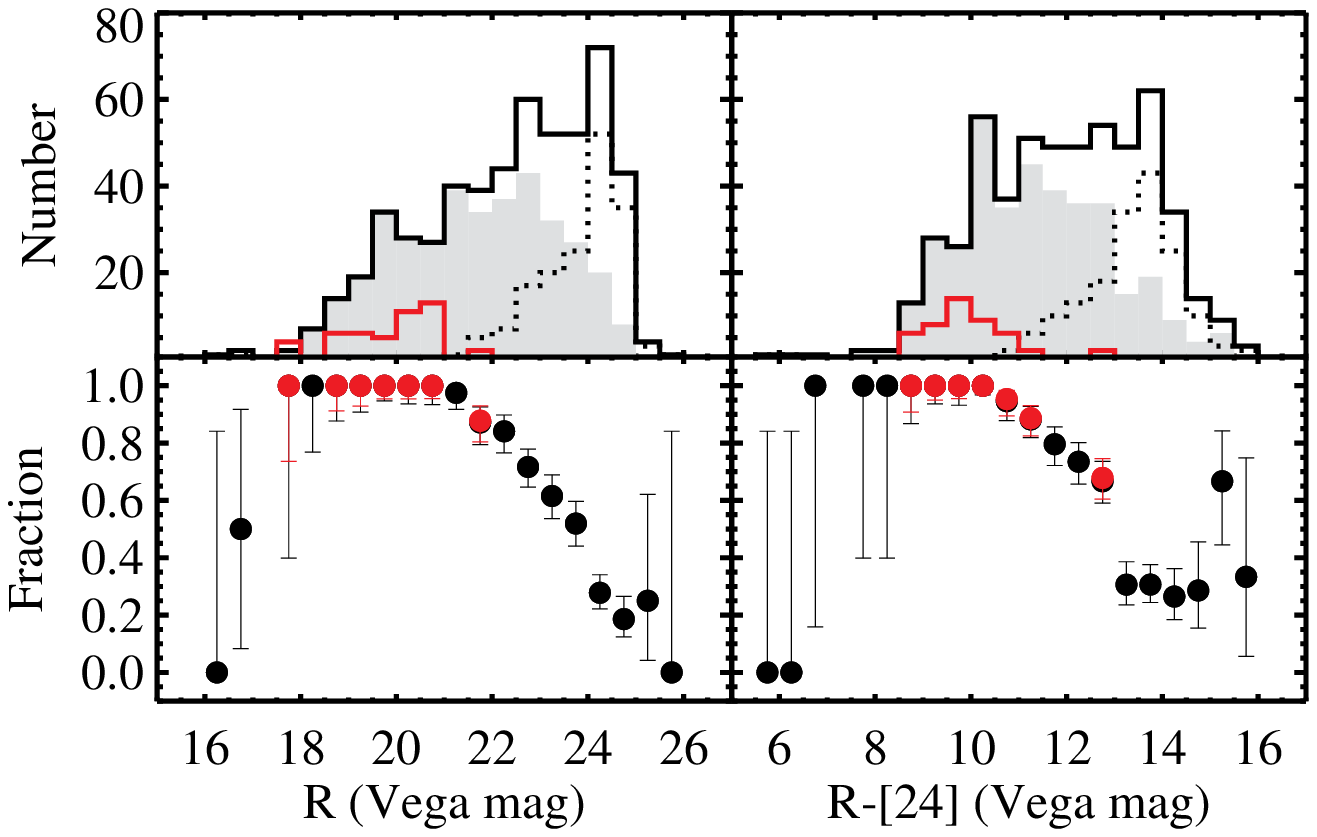}
\caption{Number and fraction of sources as a function of 24 $\micron$ flux density, $R$-band magnitude, and $R$-[24] color.  The error bars in the bottom panels include Poisson and binomial statistics \citep{Gehrels86}.  The success rate of obtaining redshifts with Keck depends strongly on $R$-band magnitude, and much less on $f_{\nu}(24 \micron)$.}
\label{fig:zrateversusf24}
\end{figure}

Figure \ref{fig:zrateversusf24} shows the success rate in determining
redshifts for Keck and Keck+AGES targets as functions of 24~$\micron$
flux density, $R$-band magnitude, and $R-[24]$ color.  This success
rate is a strong function of $R$-band magnitude, and drops below 50\%
for $R > 23.5$ mag.  This dependence on $R$-band magnitude propogates
into a dependence on the $R-[24]$ color.  In contrast, the redshift
success rate is a much weaker function of 24 $\micron$ flux density.

Figure \ref{fig:Rf24versusz} shows the 24 $\micron$ flux density,
$R$-band magnitude, and $R-[24]$ color versus redshift for the sources
with spectroscopic redshifts.  The 24 $\micron$ flux density is not
tightly correlated with redshift.  In contrast, both the $R$-band
magnitudes and the $R-[24]$ colors of the majority of these same
sources correlate strongly with redshift out to $z = 1$.  At larger
redshifts, the scatter increases substantially.

For each source with an emission line redshift, Figure
\ref{fig:lumvwave} shows the $\nu L_{\nu}$ luminosity at the rest
wavelength probed by the MIPS 24~$\micron$ filter at the observed
redshift.  The luminosity-dependent SEDs developed for the
\citet{Chary01} models are used to indicate the regions of this plot
occupied by normal galaxies, LIRGs, ULIRGs, and HyperLIRGs.  This
figure shows that the 24~$\micron$ sources in our sample span a large
range in bolometric luminosities.  

\subsection{Redshift Distribution of Galaxies without Emission Lines}
\label{sec:GalaxiesWithoutEmissionLines}

We were unable to determine emission line redshifts for 170 Keck
targets (124 DEIMOS targets and 46 LRIS targets).  Visual inspection
of the optical, IRAC, and MIPS imaging suggests that the optical
counterpart may have been incorrectly targeted in up to 20\% of these
sources.  Generally, this uncertainty is due to the faintness of the
optical counterpart combined with the large 24 $\micron$ beam size.  A
quarter of the time, the uncertainty stems from a choice between
multiple optical counterparts.  Despite this uncertainty, we use the
following algorithm to estimate the redshift distribution of these
sources.

First, we assume that these sources lie at $z < 4.5$.  This is the
highest emission-line redshift found in our survey.  This redshift
cut-off is further justified by the fact that a $z > 4.5$ source with
$f_{\nu}(24 \micron) \ge 300$ $\mu$Jy would have a luminosity of ${\rm
L}_{8-1000\micron}>3.5 \times 10^{13} \ {\rm L}_{\odot}$ (assuming the
templates of \citet{Chary01}).  Galaxies with such extreme
luminosities should be exceedingly rare.

Second, we assume that 24~$\micron$ sources that are well-detected in
the bluest NDWFS filter ($B_W < 25$ mag) must lie at $z < 3$.  This
ensures that some part of the $B_W$ filter (which covers the
approximate wavelength range 3500--4750 \AA) is sampling rest-frame
wavelengths longward of the Lyman break.

Third, we assume that all bright MIPS sources should be undergoing
strong star formation and/or AGN activity, and should therefore harbor
the conditions required to produce strong emission lines, including
Ly$\alpha$, [OII]$\lambda$3727, H$\beta$, and H$\alpha$.  Although red
sources that are bright at 24~$\micron$ must be dusty, these emission
lines are able to penetrate the dust for the majority of our targets.
This is likely due to the fact that the dust obscuration is patchy,
allowing optical/UV emission lines to be observed in very obscured
sources.  In fact, many very obscured local ultraluminous infrared
galaxies, including Arp 220, display strong emission lines which
dominate their optical spectra
\citep{Duc97,Veilleux97,Veilleux95,Cutri94,Allen91,Armus89,Kleinmann88,Sanders88b}.
We assume that this is the case for the targets without detected
emission lines as well, and that the lack of observed emission lines
is due to insufficient wavelength coverage rather than to
intrinsically weak or completely obscured emission lines.  Below we
describe how we rule out those redshifts for which any one of
Ly$\alpha$, [OII]$\lambda$3727, H$\beta$, or H$\alpha$ could have been
detected.

We assume that a line could have been detected at a particular
wavelength (corresponding to a specific redshift) if the limiting
rest-frame equivalent width at that wavelength is smaller than a
fiducial rest-frame equivalent width, which we take to be 10 \AA \ for
each of the four emission lines we are considering.  For each
spectrum, we determined the limiting rest-frame equivalent width as a
function of wavelength following Equation 4 in \citet{Hogg98}:

\begin{equation}
{\rm EQW}_{\rm lim, rest}(\lambda) = \frac{\eta \lambda_1}{r(\lambda)}\left(\frac{\Delta \lambda}{\lambda_1}\right)^{1/2}[1+z(\lambda)]^{-1/2}.
\end{equation}

\noindent Here, $\eta$ is the minimum signal-to-noise ratio required
for the line to be detected, and was taken to be three for both LRIS
and DEIMOS spectra, and for all four emission lines.  The variable
$r(\lambda)$ is the signal-to-noise ratio of the extracted spectrum as
a function of wavelength.  It was computed by convolving the extracted
spectrum with a Gaussian kernel characterized by a FWHM of 200 \AA,
and dividing the result by the noise within a moving bin of 13 pixels
in the (unconvolved) extracted spectrum.  The spectral dispersion,
$\lambda_1$, is 0.65 \AA \ pixel$^{-1}$ for DEIMOS, 1.09 \AA \
pixel$^{-1}$ for the blue side of LRIS, and 1.86 \AA \ pixel$^{-1}$
for the red side of LRIS.  The parameter $\Delta \lambda$ is the
rest-frame FWHM of the emission line in question, which we take to be
10 \AA \ for all four emission lines.  Finally, $z$ is the redshift
which corresponds to the wavelength at which we are computing the
rest-frame limiting EQW, and is given by $z = (\lambda / \lambda_{\rm
rest}) - 1$, where $\lambda_{\rm rest} = 1216, 3727, 4861, 6563$ \AA \
for Ly$\alpha$, [OII], H$\beta$, and H$\alpha$, respectively.

After ruling out those redshifts at which these emission lines could
have been detected for each individual source, we constructed a
redshift probability distribution function for the ensemble by summing
the number of sources that could lie at a given redshift, and
normalizing the result to a total probability of unity.  This is based
on the assumption that each galaxy has a uniform probability of lying
at any redshift that is not ruled out by the above three arguments
(that all 24~$\micron$ sources lie at $z < 4$, that all 24~$\micron$
sources that are well-detected in $B_W$ must lie at $z < 3$, and that
all 24~$\micron$ sources must have strong emission lines that have
gone undetected due to limited spectral coverage).  Figure
\ref{fig:nozprobdist} shows the result for all of the targets without
redshifts, and for those observed with DEIMOS and LRIS separately.
Based on the lack of the [OII]$\lambda$3727 emission line in their
spectra, most of the sources observed with DEIMOS are unlikely to lie
at $z < 1.5$.  In addition, few lie at $z > 3.5$, or else we would
have observed Ly$\alpha$.  In contrast, the redshift distribution of
the sources observed with LRIS is closer to uniform.  The similarity
of the top and bottom panels of Figure \ref{fig:nozprobdist} reflects
the fact that most of the targets without spectroscopic redshifts were
observed with DEIMOS.

Figure \ref{fig:nozprobdist} does not take into account the fact that
the mean luminosity of galaxies in a flux-limited survey increases
with redshift and the luminosity function decreases rapidly at the
high-luminosity end.  Qualitatively, this means that a galaxy does not
have an equal probability of lying at every redshift not ruled out by
the above algorithm.  A given galaxy is more likely to be an
L$_{\ast}$ galaxy at moderate redshift than five times more luminous
at a higher redshift.  However, given the limited photometric
information available to derive a luminosity at a common wavelength
for each galaxy, we do not attempt to quantify this qualitative
statement.

\subsection{Redshift Distribution}
\label{sec:RedshiftDistribution}

The redshift distribution of the Keck+AGES targets is shown in Figure
\ref{fig:rawzdist}.  The contributions of the following categories of
sources are indicated separately: the Keck targets with emission-line
redshifts; the supplementary AGES sample selected to correct the bias
of the Keck spectroscopic targeting against bright optical sources;
and Keck targets without spectroscopic redshifts, assuming that they
all have emission lines that could not be observed due to the limited
wavelength coverage of the Keck spectrographs and that they have an
equal probability of lying at any redshift in the range $0 < z < 4.5$
that is not excluded by the lack of detected emission lines (see
Section \ref{sec:GalaxiesWithoutEmissionLines}).  Figure
\ref{fig:rawzdist} shows that the galaxies with spectroscopic
redshifts peak at $z \sim 0.3$.  The majority of the targets without
emission lines cannot lie at this redshift or [OII]$\lambda$3727 would
have been observed.  Instead, most of these sources are constrained to
lie at $2 < z < 3$, where [OII]$\lambda$3727 lies longward of the
observed spectrum and Ly$\alpha$ lies shortward of it.  These sources
appear in the redshift distribution as a weak second peak at $z \sim
2$.

\section{Discussion}
\label{sec:Discussion}

\subsection{Shape of the Redshift Distribution}
\label{sec:ShapeRedshiftDistribution}

The redshift distribution of 24~$\micron$ sources is expected to be
shaped both by evolution in the infrared luminosity function and by
major features in the mid-infrared spectra of luminous infrared
galaxies.  These in turn depend upon fundamental processes governing
galaxy evolution, such as star formation, AGN activity, dust
obscuration, galaxy mergers, and feedback from supernovae and AGN.
Thus, the redshift distribution is an important constraint on models
of these processes.

In the absence of any evolution, the number of sources observed in a
given redshift bin depends only on the volume probed by that bin, the
range of luminosities probed by that bin, and the shape of the $z=0$
luminosity function.  The first two effects act in opposite
directions: volume increases with redshift, leading to an increase in
the observed number of galaxies with redshift.  However, the flux
limit means that a decreasing range of luminosities, and therefore a
decreasing number of galaxies, is observed with increasing redshift.
How does an evolving luminosity function affect the redshift
distribution?  A decrease in either the space density or the
characteristic luminosity of infrared sources with redshift would
cause fewer galaxies to be observed at high redshift.  Conversely, an
increase in the space density or characteristic luminosity would
appear as a rise in the redshift distribution.

In addition to a strong dependence on the evolving luminosity
function, the observed redshift distribution is also affected by
prominent features in the mid-infrared spectra of infrared sources.  A
strong emission feature can increase the observed broadband MIPS 24
$\micron$ flux density of a galaxy.  Thus, objects with strong
emission features can be pushed into a flux-limited sample, even if
they would not be targeted based on their continuua alone.
Conversely, sources with strong absorption features can fall out of
flux-limited samples.  The strongest mid-infrared spectral features in
LIRGs and ULIRGs are the silicate absorption feature at rest-frame 9.7
$\micron$ and the Polycyclic Aromatic Hydrocarbon (PAH) emission
features at rest-frame wavelengths 3.3, 6.2, 7.7, 8.6, 11.3, 12.7,
16.3, and 17 $\micron$.  The redshift distribution of 24~$\micron$
sources may display troughs and peaks at redshifts corresponding to
the presence of these absorption and emission features in the observed
24~$\micron$ bandpass, and therefore contains information about the
types of galaxies at each redshift.  In this subsection, we consider
the most prominent features of the observed redshift distribution
shown in Figure \ref{fig:rawzdist}, compare to the results of previous
observations, and discuss their possible origins.

Figure \ref{fig:rawzdist} shows a strong increase in the number of
sources from $z = 0$ to $z \sim 0.3$.  A rise is also seen at $z < 1$
by \citet{PerezGonzalez05}, \citet{LeFloch05}, and \citet{Caputi06}.
However, the peak in these studies occurs at higher redshifts ($z \sim
0.8$), presumably because they have a deeper 24~$\micron$ flux density
limit (83 $\mu$Jy versus our 300~$\mu$Jy). As discussed above, a rise
in the redshift distribution could be caused by a strong spectral
feature entering the 24~$\micron$ bandpass, by evolution in the
luminosity function, or simply by the fact that larger volumes are
probed at higher redshifts.  There are no strong spectral features
that could be causing the $z \sim 0.3$ peak.  The detailed modeling
that is required to distinguish between the remaining scenarios is
beyond the scope of this paper. However, in \S
\ref{sec:ComparisonWithModels} we compare our observations to existing
phenomenological models of the evolving luminosity function.
 
There is potentially another peak in the redshift distribution at $z
\sim 0.9$.  However, it is very weak, and has not been seen in the
previous measurements of the redshift distribution described in
\S{\ref{sec:Introduction}}.  Given the small areas of the previous
surveys, this could very well be due to cosmic variance (see
\citet{Somerville04}).  Additional possibilities include the fact that
most previous efforts were either incomplete at these redshifts, or
based on photometric redshifts, which would tend to smooth out
features in the intrinsic redshift distribution.  If real, this peak
could be the result of the 12.7 $\micron$ PAH feature and the 12.8
$\micron$ [NeII] emission line passing through the 24 $\micron$
bandpass.

Between $z \sim 0.8$ and $z \sim 1.6$, the redshift distribution
exhibits a steep decline, which has also been reported by
\citet{PerezGonzalez05} and \citet{Caputi06}.  Such a decline could be
the result of a flattening or a drop in either the density or the
characteristic luminosity of infrared galaxies.  A decline could also
result if a significant fraction of sources with $f_{\nu}(24 \micron)
\ge 300$ $\mu$Jy have deep silicate absorption features at rest-frame
9.7 $\micron$. Some highly absorbed sources may fall below the sample's
flux limit as the absorption feature enters the 24 $\micron$ bandpass.
This would result in a broad dip centered at $z \sim 1.5$, the
redshift at which the center of the broad absorption feature coincides
with the center of the wide bandpass.

Also visible in the redshift distribution is a possible peak at $z
\sim 2$.  However, Figure \ref{fig:rawzdist} clearly shows that this
peak is composed mostly of sources for which we were unable to
determine spectroscopic redshifts due to the lack of observed emission
lines.  We remind the reader that we constrained the redshift
distribution of such sources by assuming that 1) these non-detections
were due to the limited wavelength coverage of optical spectrographs;
2) 24~$\micron$ sources that are well-detected in the bluest NDWFS
filter ($B_W < 25$ mag) must lie at $z < 3$ so that some part of the
$B_W$ filter is sampling rest-frame wavelengths longward of the Lyman
break; and 3) there are no 24~$\micron$ sources at redshifts larger
than $z = 4.5$ (see \S \ref{sec:GalaxiesWithoutEmissionLines}).  The
exact location of the $z \sim 2$ peak is uncertain because we have
assumed that a given galaxy is equally likely to lie at any redshift
not ruled out by the previous three assumptions.  

\begin{figure}
\epsscale{0.5} 
\plotone{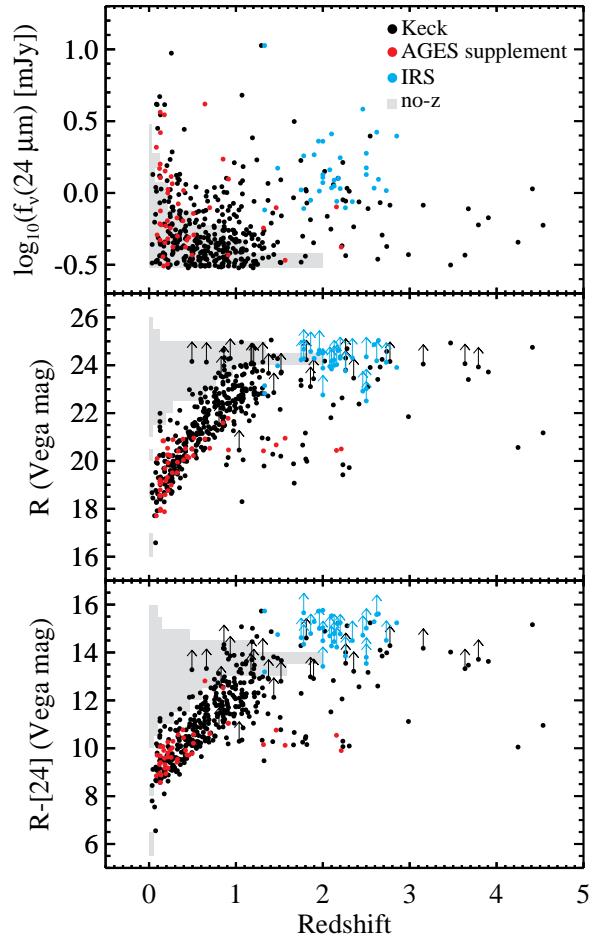}
\caption{MIPS 24~$\micron$ flux density, $R$-band magnitude, and
$R-[24]$ color versus redshift.  The points are color-coded by the source
of the redshifts.  The IRS redshifts are for a sample of \bootes \
sources (not necessarily within the Keck mask boundaries) taken from
\citet{Houck05}.  The 24 $\micron$ flux density is not strongly
correlated with redshift over any redshift range, while both the $R$-band
magnitude and the $R-[24]$ color show a correlation for $z \lesssim
1$.  However, the correlation is not tight enough and does not extend
to high-enough redshifts to predict the redshifts of the Keck targets
without spectroscopic redshifts, whose properties are indicated by the
grey histograms.}
\label{fig:Rf24versusz}
\end{figure}

\begin{figure}
\epsscale{0.9}
\plotone{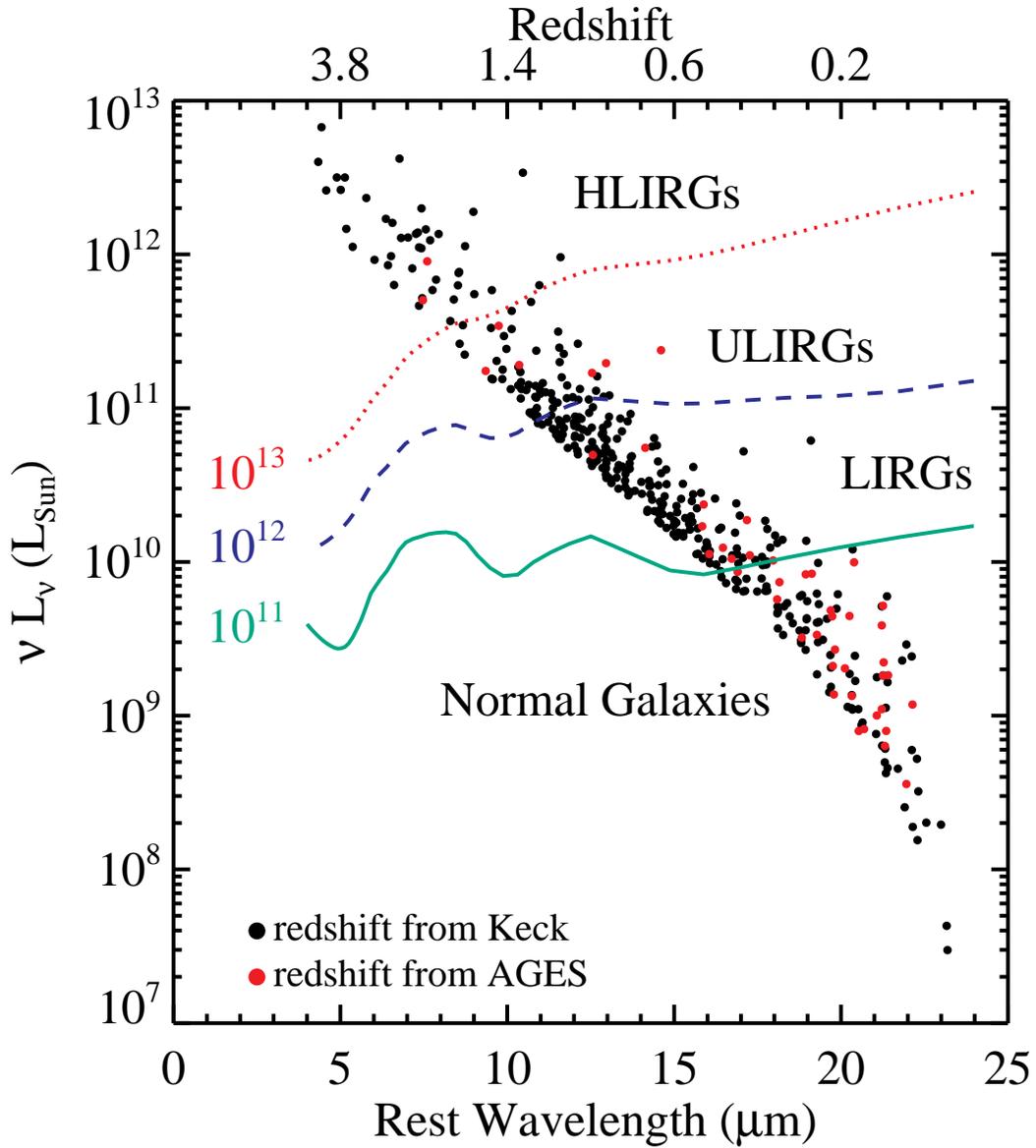}
\caption{Luminosity ($\nu L_{\nu}$) versus rest-frame wavelength for
Keck+AGES targets with spectroscopic redshifts.  Plotted quantities
were computed using observed 24~$\micron$ flux densities and emission
line redshifts.  Also shown are the tracks for template sources with
$L_{8-1000 \micron} = 10^{11}$ ${\rm L}_{\odot}$, $10^{12}$
L$_{\odot}$, and $10^{13}$ L$_{\odot}$ \citep{Chary01}.  These models
represent the boundaries between ``normal'' galaxies, luminous
infrared galaxies (LIRGs), ultraluminous infrared galaxies (ULIRGs),
and hyperluminous infrared galaxies (HLIRGs).  For sources that match
these templates the bolometric correction, i.e. L$_{8-1000 \micron} /
\nu L_{\nu}(24/(1+z) \micron)$ is $\sim$10, $\sim$10 and $\sim$30 for
LIRGs, ULIRGs and HLIRGs respectively.}
\label{fig:lumvwave}
\end{figure}

\begin{figure}
\epsscale{0.5}
\plotone{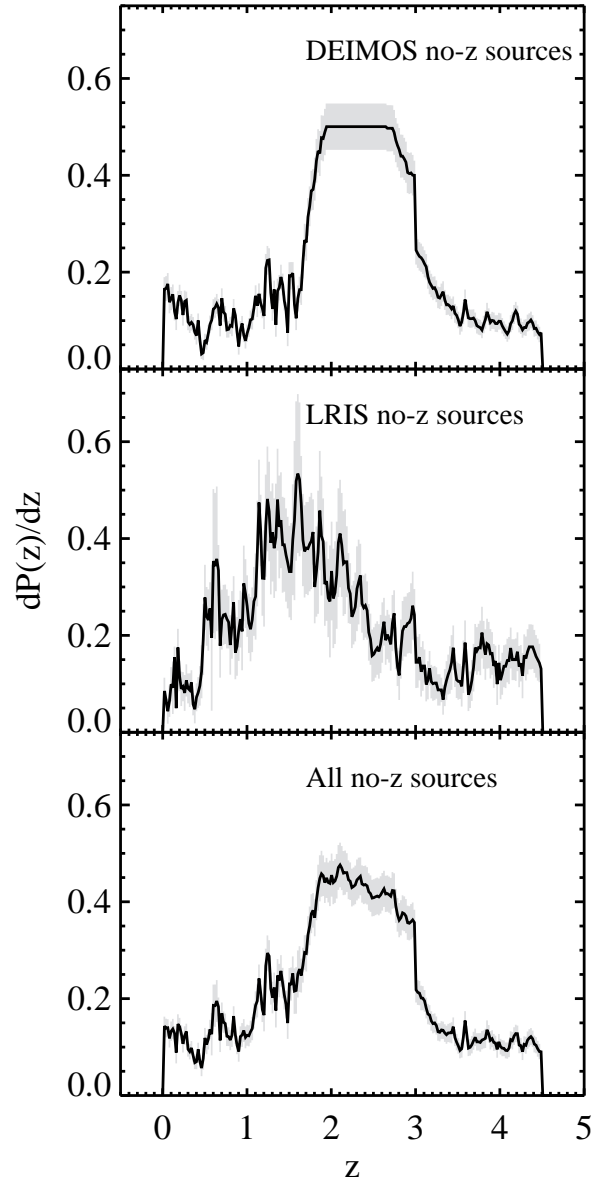}
\caption{Estimated redshift probability density function
(\textit{black line)} and Poisson errors (\textit{grey shaded region})
for Keck targets without spectroscopic redshifts.  These functions
were calculated by ruling out redshifts based on the non-detections of
Ly$\alpha$, [OII]$\lambda$3727, H$\beta$, and H$\alpha$, as described
in \S{\ref{sec:GalaxiesWithoutEmissionLines}}.  This method predicts
that the redshift distribution of the ``no-z'' sources peaks between
$1.5 < z < 3$.}
\label{fig:nozprobdist}
\end{figure}

\begin{figure}
\epsscale{0.9}
\plotone{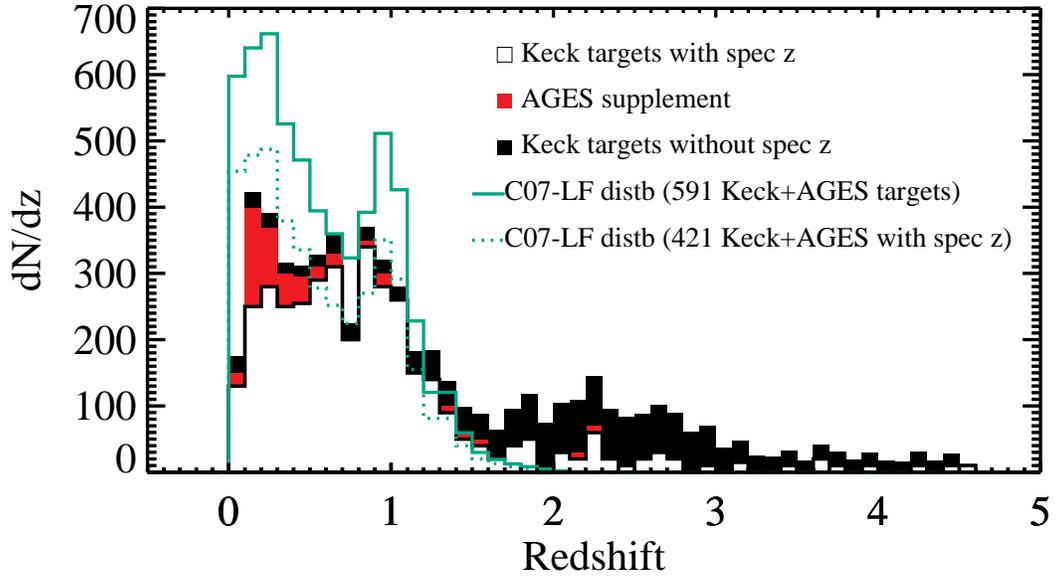}
\caption{Observed redshift distribution of 24~$\micron$ sources down
to $f_{\nu}(24) = 300$ $\mu$Jy.  The redshift distribution peaks at $z
\sim 0.3$, with a marginally-detected additional peak at $z \sim 0.9$.
We find weak evidence for another peak at $z \sim 2$, based on
estimates of the redshift distribution of the Keck targets without
spectroscopic redshifts (see
\S{\ref{sec:GalaxiesWithoutEmissionLines}}).  The redshift
distribution implied by the C07 ${\rm L}_{8-1000 \micron}$ luminosity
function is also shown, normalized to the number of sources in the
entire Keck+AGES target list, and also to the subset with
spectroscopic redshifts.  The C07 luminosity function implies that the
sources without spectroscopic redshifts lie at $z < 1$.  We argue in
\S{\ref{sec:RedshiftDistribution}} that this is unlikely.}
\label{fig:rawzdist}
\end{figure}

In the future, it will be possible to provide stronger constraints on
the redshift distribution of the mid-infrared sources which lack
spectroscopic redshifts by incorporating priors associated with the
mid-infrared luminosities, optical luminosities, and colors of the
sources.  However, adding priors to our estimate at the present time
is complicated for many reasons.  First, there is no well-measured
luminosity function for 24 $\micron$ sources at high redshift. The
current best luminosity functions are estimated using photometric
redshifts and do not extend beyond $z \sim 2$ \citep[e.g.][hereafter
C07]{Caputi07}.  Second, the sources for which we are attempting to
estimate the redshift distribution are the reddest ones whose SEDs are
poorly fit by galaxy templates and for which photometric redshifts are
the least reliable. Third, the optical luminosity distribution of 24
$\micron$ sources is not known as a function of redshift.  Although it
is expected that fainter sources will lie at higher redshift (Figure
\ref{fig:Rf24versusz}), it is not clear how to implement this
expectation and what the contribution to the counts might be, if any,
from low-redshift dwarfs. Finally, we note that our spectroscopic
redshift sample shows a clear trend of $R-[24]$ color with redshift,
but the relation is broad and it is not clear if the sources that lack
spectroscopic redshift measurements partake in the trend or are
outliers.  We made an attempt to implement a simple luminosity prior
by extrapolating the evolving luminosity function as reported by C07.
The results for all of the Keck+AGES targets as well as for the subset
of these with spectroscopic redshifts are shown in Figure
\ref{fig:rawzdist}.  Using the \citet{LeFloch05} luminosity function
extrapolated beyond $z = 1$ produces similar results.  Compared to the
C07 luminosity function, it more closely reproduces our spectroscopic
redshift distribution at $z < 1$, with the exception of the small dip
at $z \sim 0.75$.  As one might expect, a simple luminosity prior
results in weighting the redshift probability distribution of a given
source to lower redshift. This would be balanced by the optical
brightness prior, which would push faint sources to higher
redshift. Depending on the redshift evolution assumed, the strength of
the bump at $z\approx 2$ can be reduced in strength. However, since
there is no knowledge of the forms or weights of the priors discussed,
we present the redshift distribution in Figure \ref{fig:rawzdist} as
the measured distribution (white histogram) and our simple prediction
for the sources without spectroscopic redshifts as estimated from the
spectroscopic data with no priors. These should be interpreted as
likely bracketing the true redshift distribution of the 24 $\micron$
sources.

The presence of a peak at $z \sim 2$ has been difficult to establish
in previous studies of the 24~$\micron$ redshift distribution.  It was
not detected by \citet{PerezGonzalez05} but it was detected with high
significance by \citet{Caputi06}.  Uncertainties in the photometric
redshifts may be responsible for this difference.  Neither
\citet{PerezGonzalez05} nor \citet{Caputi06} have very many ($\la$25)
spectroscopic redshifts above $z = 1.5$ (See Figures 13 and 14 in the
former and Figures 1 and 5 in the latter).  As a result, the scatter
in the photometric redshifts at $z > 1.5$ has not been quantified.
Without this information, an assessment of the reality of a peak at $z
\sim 2$ based on photometric redshifts is difficult.  (For comparison,
our Keck spectroscopic survey includes spectroscopic redshifts for 46
galaxies at $z > 1.5$.)  Both P05 and C05 focused on sources down to
83 $\mu$Jy.  Our study suggests that even sources brighter than 300
$\mu$Jy display this peak, although this is based on assumptions
regarding the nature of the sources without detected emission lines.

Neural net and template-fitting techniques have been used to calculate
photometric redshifts \citep{Brodwin06} for NDWFS sources detected by
IRAC, including the sources presented in this work.  These photometric
redshifts are highly accurate for IRAC-selected galaxies and AGN out
to $z = 1.5$ and $z = 3$, respectively.  An analysis of 24~$\micron$
selected galaxies with spectroscopic redshifts shows that their
photometric redshifts are accurate out to $z \sim 1$, where the 24
$\micron$ sample is is dominated by normal galaxies and LIRGs.  At
higher redshifts, corresponding to ULIRGs, the photometric redshifts
become unreliable (the redshift dispersion in $\sigma/(1+z)$ is 0.13
with 95\% clipping).  Since it is likely that the Keck sources without
spectroscopic redshifts lie at $z > 1$, we cannot use these
photometric redshifts to determine their redshift distribution.

There are two reasonable interpretations of the $z \sim 2$ peak in the
redshift distribution.  The first is that there is a significant
population of infrared-luminous sources with strong 7.7 and 8.6
$\micron$ PAH features.  These sources have increasing detectability
at $1.6 < z < 2$ and decreasing detectability at $2 < z < 3$ as the
PAH emission shifts in and out of the 24~$\micron$ bandpass.  The
sources comprising the $z \sim 2$ peak have $f_{\nu}(24 \micron) \ge
300$ $\mu$Jy, and are therefore ULIRGs with infrared luminosities
L$_{8-1000 \micron} > 3 \times 10^{12}$ L$_{\odot}$ (assuming the
templates of \citet{Chary01}).  In the local universe, galaxies with
these luminosities typically have low PAH equivalent widths
\citep{Lutz98,Tran01,Desai07b}.  However, there are sources at $z \sim
2$ that are known to be both extremely luminous and have strong PAH
features \citep{Lutz05,Yan05,MenendezDelmestre07,Sajina07}, indicating
that the relationship between PAH equivalent width and luminosity may
be evolving in the sense that extremely luminous galaxies at high
redshift have spectral properties similar to less luminous galaxies at
low redshift \citep{Sajina07,Desai07b}.  Such an evolution might be
expected if the PAH equivalent width is a function of various evolving
sources properties (e.g.~the distribution of star formation and the
presence of an AGN).  If the relationship between luminosity and
spectral properties does evolve, then it is reasonable that a
significant fraction of the sources composing the $z \sim 2$ peak have
large PAH equivalent widths.  This is an important effect to include
in phenomenological models of galaxy evolution, such as those
discussed in the next subsection.

A second explanation for the shape of the redshift distribution at $z
> 1.6$ is that there is a large population of sources at these
redshifts with deep 9.7 $\micron$ silicate absorption features.  As
discussed above, similar sources could be at least partly responsible
for the decline in the redshift distribution between $z \sim 0.8$ and
$z \sim 1.6$ because of decreasing detectability as the 24~$\micron$
bandpass samples the absorption feature.  In contrast, these sources
are more likely to be detected at $z \sim 2$, where the 24~$\micron$
bandpass is sampling rest-frame wavelengths shortward of the
absorption feature.  In this scenario, the $z \sim 2$ peak represents
the reappearance of highly absorbed sources as the silicate absorption
feature moves out of the 24~$\micron$ bandpass.  The decrease at $z >
2$ would then be due to a combination of a decline in the flux density
at wavelengths shorter than 8 $\micron$ and the fact that fewer
galaxies meet the flux density cut at higher redshift in the absence
of evolution.  

We have qualitatively argued that the $z \sim 2$ bump could consist of
some combination of sources with large PAH equivalent widths and
sources with power-law continuua accompanied by significant silicate
absorption.  Is there any evidence that either of these populations
dominates?  Figure \ref{fig:rawzdist} shows that a small fraction of
the sources in the $z \sim 2$ peak have been spectroscopically
confirmed.  Of the 33 Keck targets with $1.6 < z < 3$, 73\% show
spectroscopic signatures of AGN activity (lines broader than 2000 km
s$^{-1}$ in the rest frame or the presence of high-ionization lines
such as NV$\lambda$1240, CIV$\lambda$1549, HeII$\lambda$1640,
[NeV]$\lambda$3346, or [NeV]$\lambda$3426).  The remaining 27\% are of
unknown type.  Of the AGN, 50\% are Type 1 AGN (with lines broader
than 2000 km s$^{-1}$).  Thus, a significant fraction of the
spectroscopically confirmed sources in the $z \sim 2$ peak are AGN.
However, these represent a small fraction of all of the galaxies in
the peak, leaving the nature of the majority unknown.

\begin{figure}
\epsscale{1}
\plotone{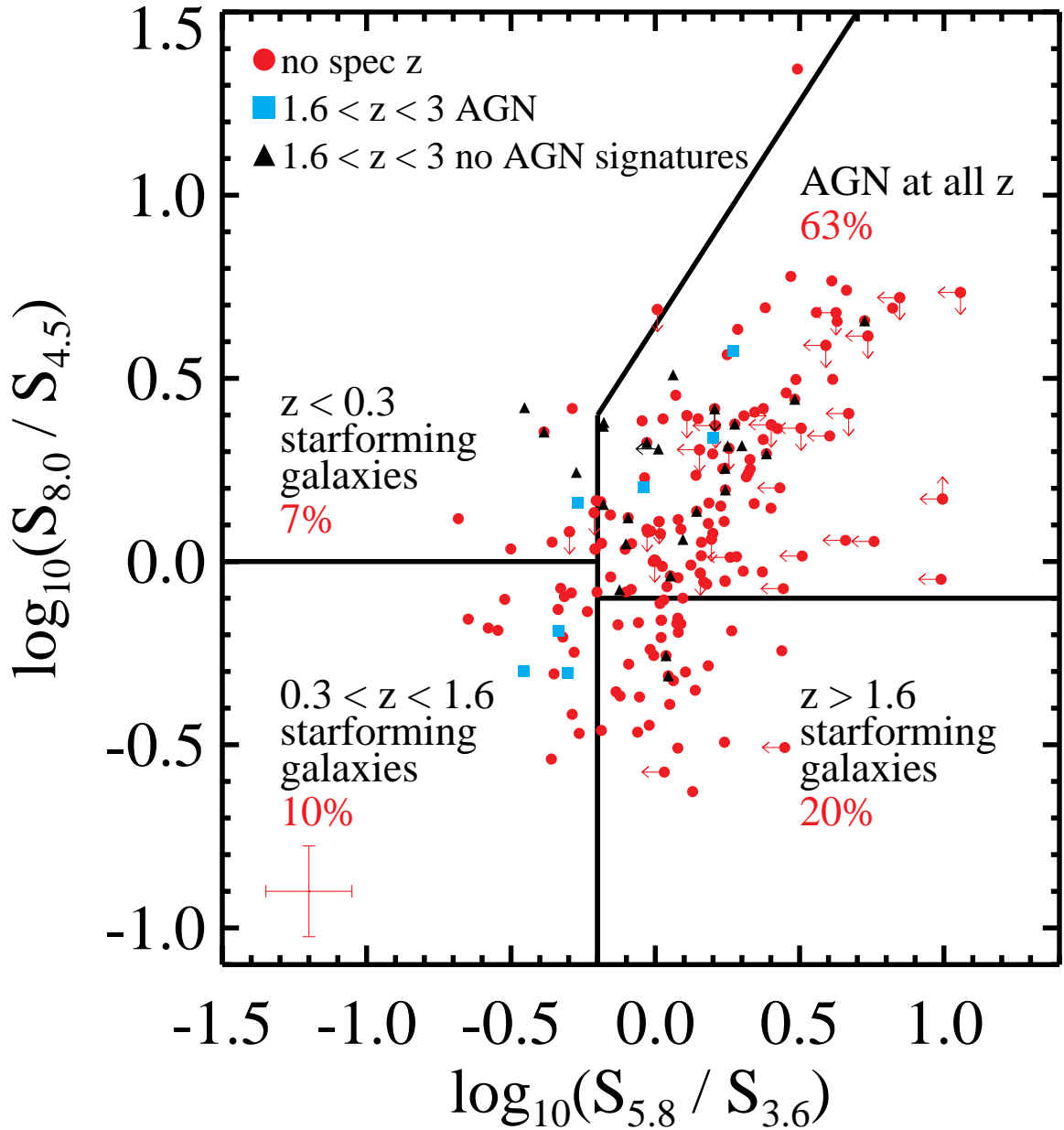}
\caption{IRAC color-color diagram of Keck targets without
spectroscopic redshifts and of Keck targets spectroscopically
confirmed to lie within the $z \sim 2$ peak.  The latter are
color-coded by whether or not they show optical spectroscopic
signatures of AGN activity.  The error bars in the bottom left of the
plot indicate the median errors for the plotted data.  The black lines
delineate the regions expected to be occupied by AGN and starbursts at
various redshifts, according to the models of \citet{Sajina05}.  The
percentage labelled in each region indicate the percentage of red
points that fall within them.  We conclude that 55\% of the sources
within the $z \sim 2$ peak are AGN-dominated (see
\S{\ref{sec:ShapeRedshiftDistribution}}).}
\label{fig:irac_keck}
\end{figure}

IRAC colors have been used by several authors to identify sources that
have power-law SEDs and therefore presumably contain powerful AGN
\citep{Stern05,Lacy04,Lacy07}.  Using the photometry from the IRAC
Shallow Survey of the NDWFS \bootes \ field \citep{Eisenhardt04},
Figure \ref{fig:irac_keck} shows the IRAC color-color diagram for all
of the Keck targets without spectroscopic redshifts. \citet{Sajina05}
have modelled the IRAC color-color diagram, and have delineated
regions expected to be occupied by starbursts and AGN at various
redshifts.  Given the uncertainties involved, the redshift
distribution implied by the IRAC colors is comparable to that deduced
from the spectral coverage of the spectra.  For example, the fraction
of starforming galaxies at $z < 1.6$ suggested by the IRAC colors is
similar to the total number of ``no-z'' sources expected to lie at $z
< 1.6$ according to Figure \ref{fig:nozprobdist}.

Given this concurrence, we assume that the vast majority of sources in
the IRAC AGN wedge lie at $z > 1.6$.  According to Figure
\ref{fig:nozprobdist}, 75\% of sources at $z > 1.6$ are expected to
lie in the $z \sim 2$ peak, which we define to include all galaxies in
the range $1.6 < z < 3$.  If the fraction of AGN-dominated galaxies at
$z > 1.6$ reflects the fraction within the $z \sim 2$ peak, then about
63\% $\times$ 75\% = 47\% of sources without spectroscopic redshifts
that lie within the $z \sim 2$ peak (50 out of 106 sources) are
AGN-dominated.

We have already mentioned that 73\% of the 33 spectroscopically
confirmed sources within the $z \sim 2$ peak show optical
spectroscopic signatures of AGN activity.  Figure \ref{fig:irac_keck}
shows that 85\% of these fall within the IRAC AGN wedge.  In fact,
82\% (27 of 33) of all of the spectroscopically confirmed sources
within the $z \sim 2$ peak lie within the AGN wedge.  Only two AGES
sources lie within the $z \sim 2$ peak (see Figure
\ref{fig:rawzdist}).

Based on the proportion of sources within the AGN wedge, we estimate
that (50 + 27) / (106 + 33) = 55\% of sources (including those with
and without detected emission lines) within the $z \sim 2$ peak are
AGN-dominated.  As mentioned in Section 1, Caputi et al. (2006) have
used photometric redshifts to estimate the redshift distribution of 24
$\micron$ sources down to 80 $\mu$Jy.  As in the current work, they
also detect a peak at $z \sim 2$, which they interpret as the result
of a population of starburst-dominated sources.  This is not
necessarily in contradiction with our results, since their flux limit
is significantly fainter than ours.  As shown in Brand et al. 2006 and
Dey et al. 2008, the fraction of sources dominated by star formation
is expected to increase with decreasing 24 $\micron$ flux
density. Indeed, Figure 5 in Caputi et al. 2006 shows that most of the
sources in their $z \sim 2$ peak fall below the 24 $\micron$ flux
density limit of our survey.

The Infrared Spectrograph \citep[IRS;][]{Houck04} on board the {\em
Spitzer Space Telescope} has been used to obtain mid-infrared spectra
of a sizeable number of optically faint ($R > 23$ mag), infrared
bright ($f_{\nu}(24 \micron) > 750$ $\mu$Jy) sources
\citep{Houck05,Yan05,Weedman06,Sajina07,Dey08}.  The IRS spectra of
such sources tend to have power-law continuua indicative of AGN
activity, and some of these also exhibit silicate absorption features
implying $z \sim 2$.  Near-infrared spectra of a similar sample also
reveal high redshifts and rest-frame optical spectroscopic signatures
of AGN activity \citep{Brand07}.  At least some of the sources in the
$z \sim 2$ peak have similar $R$-band magnitudes and 24~$\micron$ flux
densities as these samples (see shaded histograms in Figure
\ref{fig:Rf24versusz}).  However, many of the sources in the $z \sim
2$ peak are substantially fainter.  There is some evidence that the
contribution of star formation to the bolometric luminosity of a
galaxy increases with decreasing 24~$\micron$ flux density
\citep{Brand06,Dey08}.  If this is the case, then it is possible that
the fainter sources in the $z \sim 2$ peak display strong PAH
emission, which is typical in starburst galaxies.  Near-infrared or
mid-infrared IRS spectroscopy of the Keck targets without redshifts
would help clarify the presence of the $z \sim 2$ peak and the energy
generation mechanisms of the sources that comprise it.

\subsection{Comparison with Models}
\label{sec:ComparisonWithModels}

In the previous subsection, we qualitatively discussed the possible
origins of the observed features of the redshift distribution shown in
Figure \ref{fig:rawzdist}.  We found that quantitative models are
required for a robust interpretation of these features.  In this
subsection, we compare our observations to the predictions of several
such models which have been found in the literature.

\begin{figure}
\epsscale{0.9}
\plotone{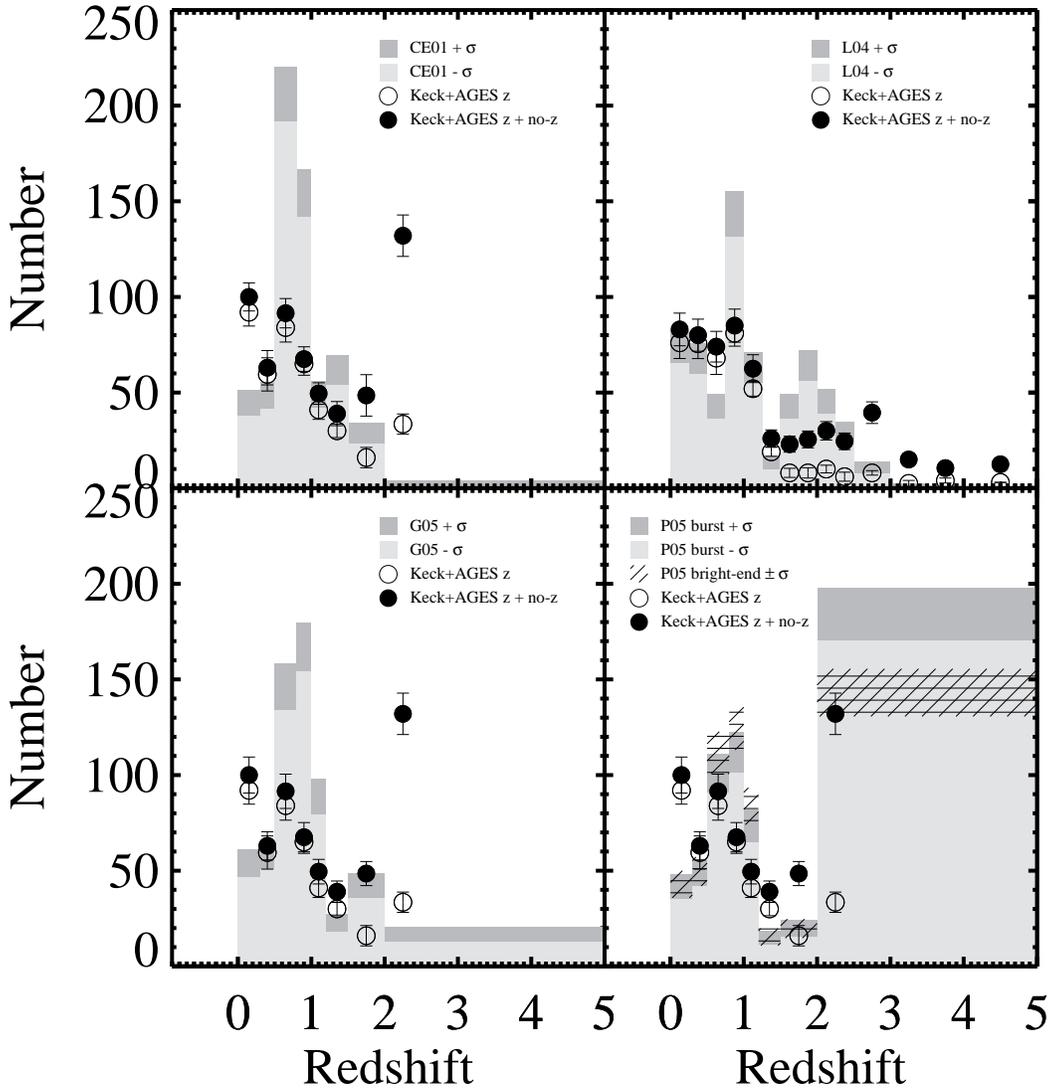}
\caption{Observed redshift distribution of 24~$\micron$ sources
compared to the CE01, L04, G05, and P05 models.  The models and their
associated Poisson errors are shown as shaded histograms.  P05 present
two variants of their model: ``burst'' and ``bright end''.  All models
have been normalized to the total number of Keck+AGES targets.  The
highest-redshift bin shows the number of counts at redshifts greater
than its leftmost extent. The observed redshift distribution, binned
identically to the models, is shown as points with 1$\sigma$ Poisson
error bars.  None of the error bars include any contribution from
cosmic variance.  The L04 and P05 models, both of which predict a
large fraction of PAH-rich ULIRGs at $z > 1.5$, best reproduce the
observations.}
\label{fig:modelcomparisonnoz}
\end{figure}

In Figure \ref{fig:modelcomparisonnoz}, we compare the observed
redshift distribution to that predicted by four models of the
evolution of mid-infrared sources.  The models are described in detail
by \citet{Chary01}, \citet{Lagache04}, \citet{Gruppioni05}, and
\citet{Pearson05}; hereafter referred to as CE01, L04, G05, and P05,
respectively.  Below, we briefly describe the major elements of these
models.

CE01 evolve the local mid-infrared luminosity function to fit the
spectrum of the cosmic infrared background and the galaxy counts at
mid-infrared, far-infrared, and submillimeter wavelengths.  They
constructed a family of luminosity-dependent template SEDs spanning
the luminosities of normal galaxies, starbursts, LIRGs, and ULIRGs.
In their model, all galaxies with $L_{8-1000 \micron} > 10^{10.2} \
{\rm L}_{\odot}$ evolve, while only 5\% of galaxies with lower
luminosities evolve.  The luminosity evolution goes as $(1+z)^{4.5}$
up to $z = 0.8$, after which it remains constant.  The density
evolution goes as $(1+z)^{1.5}$ out to $z = 0.8$, after which it goes
as $(1+z)^{-0.4}$.

L04 follow the evolution of ``normal'' and starburst galaxies.  Each
of these galaxy types is represented by luminosity-dependent SEDs.
The magnitude of the luminosity and density evolution in these
populations was chosen to fit the existing observed mid- and
far-infrared number counts, the mid- and far-infrared redshift
distributions, the far-infrared luminosity functions, the far-infrared
background, and its fluctuations.  The resulting model features a very
high rate of evolution of the luminosity density of starburst
galaxies, peaking at $z \sim 0.7$ and remaining constant up to $z =
4$.  The normal galaxies evolve out to $z = 0.4$, after which their
luminosity density remains constant.

G05 start with the 15 $\micron$ luminosity function of galaxies and
AGN.  They assume that four populations contribute: ``normal''
galaxies, starburst galaxies, Type 1 AGN, and Type 2 AGN.  They
constrain the luminosity and density evolution of these populations by
requiring a match to the observed 15 $\micron$ number counts.  In this
model, the normal population does not evolve.  Meanwhile, the
starburst population evolves strongly out to $z = 1$, as $(1+z)^{3.5}$
in luminosity and as $(1+z)^{3.8}$ in density.  Type I AGN evolve in
luminosity as $(1+z)^{2.6}$ out to $z = 2$, with no evolution at
higher redshifts.  Type II AGN evolve in luminosity as rapidly as
$(1+z)^{2.6}$ out to $z = 2$ and do not evolve at higher redshifts.

P05 present two phenomenological models tracking the evolution of four
classes of galaxies: normal, starburst, LIRG/ULIRG, and AGN.  Both
models are anchored to observed, type-dependent, local mid-infrared
luminosity functions.  In the ``bright-end'' model, the starbursts and
AGN evolve in both luminosity and density as $(1+z)^{3.3}$ out to $z =
1$, and do not evolve at higher redshifts.  The LIRG/ULIRG component
evolves in luminosity as $(1+z)^{2.5}$ and in density as $(1+z)^{3.5}$
out to $z = 1$, and does not evolve at higher redshifts.  In the
``burst'' model, the starburst population behaves as it did in the
``bright-end'' model.  However, the LIRG/ULIRG component evolves in
density as $(1+z)^7$ out to $z = 1$ and then more slowly in both
density and luminosity to higher redshifts.

For a fair comparison with the observations, all of the models shown
in Figure \ref{fig:modelcomparisonnoz} have been computed down to
$f_{\nu}(24 \micron) = 300$ $\mu$Jy and have been normalized to the
number of 24~$\micron$ sources targeted by our Keck observations
(591).  The observed redshift distributions shown in Figure
\ref{fig:modelcomparisonnoz} have been binned in redshift identically
to the models for easy comparison.  Because the models all have
different binsizes, which are also different from the binsize used in
Figure \ref{fig:rawzdist}, the observations look slightly different in
each panel and in Figure \ref{fig:rawzdist}.

In comparing the models to the observations in Figure
\ref{fig:modelcomparisonnoz}, we note that the detailed structure of
the observed redshift distribution at $z > 1.5$ is not
well-constrained, for the reasons described in
\S{\ref{sec:GalaxiesWithoutEmissionLines}}.  However, the relative
number of sources at $z < 1$ and $z > 1$ should be a robust statistic
with which to compare models and observations, since the vast majority
of the Keck sources without spectroscopic redshifts are extremely
unlikely to lie at $z < 1$.  The observations indicate that 45\% of
the Keck+AGES targets lie at $z > 1$.  In the CE01 and G05 models,
only $\sim$25\% of the sources lie at $z > 1$.  The templates used in
these models may be insufficient to reproduce the observations.  The
CE01 model lacks both highly absorbed AGN-dominated sources and
high-luminosity sources with strong PAH features, both of which could
induce a peak at $z \sim 2$, as discussed in
\S{\ref{sec:ShapeRedshiftDistribution}}.  The G05 model includes no
dependence of PAH strength on luminosity.  Although they include
AGN-dominated sources, none have deep silicate absorption.

In contrast, the L04 and P05 models correctly predict the fraction of
sources at $z > 1$.  In both models, these high-redshift sources are
starburst-dominated ULIRGs with strong PAH features.  However, we
suggested in \S{\ref{sec:ShapeRedshiftDistribution}} that absorbed
power-law (AGN-dominated) SEDs likely also contribute to a trough at
$z \sim 1.5$ which appears as a peak at $z \sim 2$.  Although the
relative numbers of $z < 1$ and $z > 1$ galaxies in the L04 and P05
models are consistent with our observations, further modelling is
needed to determine whether an equally consistent model could be built
with a higher fraction of AGN-dominated sources at $z > 1$.  In
addition, the exact position of the $z \sim 2$ peak is critical to
testing the L04 and P05 models.  This would require additional
spectroscopy in the near- or mid-infrared.

\section{Conclusions}
\label{sec:Conclusions}

We have conducted a redshift survey of 591 bright ($f_{\nu}(24
\micron) \ge 300$ $\mu$Jy) MIPS sources in the NDWFS \bootes \ field.
We obtained emission line redshifts for 71\% of our targets, and have
developed an algorithm to derive the redshift distribution of the
remaining 29\%.  Using the entire distribution, we conclude:

\begin{enumerate}

\item The redshift distribution of bright 24~$\micron$ sources peaks
at $z \sim 0.3$.  There are no strong spectral features entering the
24~$\micron$ bandpass at the appropriate redshifts to cause this peak.
It is therefore likely caused either by evolution in the luminosity
function or by the fact that larger volumes are probed at higher
redshifts, possibilities which can be tested with models.

\item There is a marginally-detected additional peak at $z \sim 0.9$.
If real, it could be due to the fact that 24~$\micron$ surveys are
more sensitive at this redshift to LIRGs with strong 12.7~$\micron$
PAH features and/or 12.8~$\micron$ [NeII] line emission.  Additional
data are needed to test the validity of this peak.  The cleanest
method would be to continue our spectroscopic survey of 24 $\micron$
sources to build up the sample size.  Another approach would be to
compute photometric redshifts for all of the 24 $\micron$ sources in
the \bootes \ field, trading redshift accuracy (reasonable for
photometric redshifts less than $z \sim 1$) for a larger sample size.

\item We find weak evidence for another peak at $z \sim 2$, consisting
mainly of sources for which we were unable to determine emission-line
redshifts, but whose redshift distribution we have estimated by
assuming that the lack of emission lines is due to the limited
wavelength coverage of the Keck spectrographs.  Such a peak would be
expected due to the increased sensitivity to starforming ULIRGs at
redshifts where their 7.7 and 8.6~$\micron$ PAH features pass through
the 24~$\micron$ bandpass.  Alternatively, the peak could be the
result of decreased sensitivity to heavily obscured AGN at redshifts
where their 9.7~$\micron$ silicate absorption features pass through
the 24~$\micron$ filter.  Based on the arguments presented in the next
paragraph, we suggest that the latter is an important effect.

\item We suggest that 55\% of the sources in the $z \sim 2$ peak are
AGN-dominated, and the remaining 45\% are starburst-dominated.  This
suggestion is based mainly on the diversity of IRAC colors among the
Keck targets without redshifts.  In addition, AGN signatures are
present in the optical spectra of 73\% of the few (33)
spectroscopically-confirmed sources within the $z \sim 2$ peak.
Finally, sources similar to the brightest ($f_{\nu}(24 \micron) > 750$
$\mu$Jy) Keck targets without spectroscopic redshifts have AGN-like
near- and mid-infrared spectra.

\item Of the four existing phenomenological models of galaxy formation
that we considered, those that include a significant number of
PAH-rich (and therefore presumably star-formation dominated) ULIRGs at
$z > 1.5$ best reproduce the relative number of galaxies observed at
$z < 1$ and $z > 1$.  However, it remains to be shown that models
including a significant fraction of AGN-dominated ULIRGs at $z > 1$
cannot also fit the observations.

\end{enumerate}

\acknowledgments
\label{sec:acknowledgments}

We appreciate the referee's thoughtful feedback, which led to
significant improvements in the paper.  We thank Ranga-Ram Chary,
David Elbaz, Carlotta Gruppioni, Guilaine Lagache, and Chris Pearson
for providing the predictions of their models.  We thank Chris
Kochanek, Daniel Eisentein, and the AGES Team for providing access to
the AGES database.  AD thanks the Spitzer Science Center for their
hospitality while this paper was being written.  This work is based in
part on observations made with the \textit{Spitzer Space Telescope},
which is operated by the Jet Propulsion Laboratory, California
Institute of Technology under a contract with NASA. Support for this
work was provided by NASA through an award issued by JPL/Caltech.  The
\textit{Spitzer} MIPS survey of the \bootes \ region was obtained
using GTO time provided by the \textit{Spitzer} Infrared Spectrograph
Team (PI: James Houck) and by M.~Rieke. This work made use of images
and data products provided by the NOAO Deep Wide-Field Survey
\citep{Jannuzi99}, which is supported by the National Optical
Astronomy Observatory (NOAO). NOAO is operated by AURA, Inc., under a
cooperative agreement with the National Science Foundation.  The
authors wish to recognize and acknowledge the very significant
cultural role and reverence that the summit of Mauna Kea has always
had within the indigenous Hawaiian community.  We are most fortunate
to have the opportunity to conduct observations from this mountain.

{\it Facilities:} \facility{Keck:I (LRIS)}, \facility{Keck:II
(DEIMOS)}, \facility{MMT (Hectospec)}, \facility{Mayall (MOSAIC-I)},
\facility{Spitzer (MIPS, IRAC, IRS)}

\bibliographystyle{apj}
\bibliography{references}

\clearpage

\end{document}